\documentclass[twocolumn,resetfootnote]{aastex7_preprint} % use for arXiv submission.

\usepackage{verbatim}
\usepackage{esvect}
\usepackage{bm}
\usepackage{amsmath}
\usepackage{multirow}
\usepackage{comment}

\makeatletter
\let\oldthebibliography\thebibliography
\let\endoldthebibliography\endthebibliography
\renewenvironment{thebibliography}[1]{%
  \oldthebibliography{#1}%
  \footnotesize % change bibliography font size
}{%
  \endoldthebibliography
}
\makeatother

\newcommand{\kms}{km~s$^{-1}$}

\newcommand{\cmg}{cm$^2$~g$^{-1}$}
\newcommand{\Msolar}{M$_{\rm \odot}$}
\newcommand{\sm}{$\sigma/m$}
\newcommand{\mdm}{$m_{\rm \scriptscriptstyle DM}$}

\usepackage{xcolor}
\definecolor{darkgreen}{rgb}{0,0.7,0}
\definecolor{mypurple}{rgb}{0.7,0,0.5}

\makeatletter
\let\frontmatter@title@above=\relax
\makeatother

\begin{document}
\rightline{FERMILAB-PUB-26-0348-T}

\title{ \vspace{0.3in} Mergers Matter: Gravothermal Collapse in Dwarf Halos with Self-Interacting Dark Matter}

\author[orcid=0000-0002-4965-0907,sname='Silverman']{Maya Silverman}
\affiliation{Department of Physics and Astronomy, University of California, Irvine, Irvine, CA 92697, USA}
\affiliation{DARK, Niels Bohr Institute, University of Copenhagen, Copenhagen 2200, Denmark}
\email[show]{maya.silverman@nbi.ku.dk}

\author[orcid=0000-0002-5560-8668, sname='Hussein']{Abdelaziz Hussein}
\affiliation{Department of Physics and MIT Kavli Institute for Astrophysics and Space Research, Massachusetts Institute of Technology, Cambridge, MA 02139, USA}
\email{abdelh@mit.edu}

\author[orcid=0000-0002-8354-7356, sname='Arora']{Arpit Arora}
\affiliation{Department of Astronomy and DiRAC Institute, University of Washington, 3910 15th Ave NE, Seattle, WA, 98195, USA}
\email{arora125@sas.upenn.edu}

\author[orcid=0000-0002-8495-8659, sname='Lisanti']{Mariangela Lisanti}
\affiliation{Center for Computational Astrophysics, Flatiron Institute, New York, NY 10010, USA}
\affiliation{Department of Physics, Princeton University, Princeton, NJ 08544, USA}
\email{mlisanti@princeton.edu}

\author[orcid=0000-0001-8555-0164, sname='Kaplinghat']{Manoj Kaplinghat} 
% \altaffiliation{Las Campanas Observatory}
\affiliation{Department of Physics and Astronomy, University of California, Irvine, Irvine, CA 92697, USA}
\email{mkapling@uci.edu}

\author[orcid=0000-0003-2806-1414, sname='Necib']{Lina Necib}
\affiliation{Department of Physics and MIT Kavli Institute for Astrophysics and Space Research, Massachusetts Institute of Technology, Cambridge, MA 02139, USA}
\email{lnecib@mit.edu}

\author[orcid=0009-0005-1250-1800, sname='Thoyas']{Andreas Thoyas}
\affiliation{Northeastern University, Boston, MA 02115, USA}
\email{athoyas@mit.edu}

\author[orcid=0000-0002-7968-2088, sname='ONeil']{Stephanie O'Neil}
\affiliation{Department of Physics and Astronomy, University of Pennsylvania, Philadelphia, PA 19104, USA}
\affiliation{Department of Physics, Princeton University, Princeton, NJ 08544, USA}
\email{sloneil@sas.upenn.edu}

\author[orcid=0000-0003-3939-3297, sname='Sanderson']{Robyn E. Sanderson}
\affiliation{Department of Physics and Astronomy, University of Pennsylvania, Philadelphia, PA 19104, USA}
\email{robynes@sas.upenn.edu}

\author[orcid=0000-0002-6196-823X, sname='Shen']{Xuejian Shen}
\affiliation{Department of Physics and MIT Kavli Institute for Astrophysics and Space Research, Massachusetts Institute of Technology, Cambridge, MA 02139, USA}
\email{xuejian@mit.edu}

\author[orcid=0000-0002-3430-3232]{Jorge Moreno}
\affiliation{Department of Physics \& Astronomy, Pomona College, Claremont, CA 91711, USA}
\affiliation{Carnegie Observatories, 813 Santa Barbara st, Pasadena, CA 91101, USA}
\email{jorge.morenosoto@pomona.edu}

\begin{abstract}

   Self-Interacting Dark Matter (SIDM) models with large cross sections at relative velocities below $\sim 100$~\kms{} can be tested with dwarf galaxy observations. We analyze six dark-matter-only zoom-in $\sim10^{10}\,{\rm M}_\odot$ halos with diverse assembly histories, adopting a cross section over mass of $\sigma/m = 70$~\cmg. We find that mergers inject orbital kinetic energy into the halo, altering the heat transport and the gravothermal evolution of the core. Three of the six halos---those with the most quiescent merger histories---show clear signs of core collapse in these simulations. Halos with sustained mergers do not collapse. Furthermore, merger-induced heat transport drives two non-collapsing halos to central densities well below the predictions of the gravothermal fluid model.  These findings suggest a novel mechanism for producing dark-matter-deficient galaxies and expanding the diversity of rotation curves beyond what halo concentration alone predicts. Merger histories are thus essential for understanding central density distributions of dwarf galaxy halos in SIDM.

\end{abstract}

% \keywords{Dark matter~(353) --- Galaxy dark matter halos~(1880) --- N-body simulations~(1083) --- Dwarf galaxies~(416)}

\section{Introduction}
\label{sec:intro}

Self-Interacting Dark Matter~(SIDM) is a promising alternative to the collisionless Cold Dark Matter~(CDM) paradigm that can modify the internal structure of halos while preserving the success of large-scale structure predictions~\citep[for reviews, see][]{Tulin:2018,Adhikari:2022sbh}.
For this reason, SIDM has been proposed to explain gaps between CDM predictions and observations on galactic scales like the prevalence of constant-density dark-matter~(DM) cores~\citep{Spergel:1999mh, Firmani2000, burkert2000structure, Rocha:2012jg, Vogelsberger:2012ku, Kaplinghat:2015aga,Elbert:2018ApJ}.
% in , including the core--cusp, missing satellites, diversity, and too-big-to-fail problems
Beyond astrophysical motivations, DM self interactions arise naturally in many dark sector models~\citep[e.g.,][]{McDonald_2002, Feng_2010}, providing independent motivation from particle physics.

In SIDM, particle scatterings enable efficient heat transport in a halo. Scattering initially causes thermalization in the halo center, leading to a uniform-density isothermal core~\citep{Rocha:2012jg,Elbert15}.
Once the core gets hotter than the outer parts, the direction of heat flow reverses, and the core becomes hotter as it loses energy to the outer halo due to the negative heat capacity of self-gravitating systems.  This steepens the temperature gradient, which drives an increases in the central core density and a contraction in the core size---a runaway process that results in gravothermal collapse, also referred to as core collapse/gravothermal catastrophe~\citep{Zel'dovich1966, lynden-bell_gravo-thermal_1968, larson_method_1970, Lightman1978, Shapiro1985, hachisu_gravothermal_1978, lynden-bell_1980, balberg-SMBH-2002, Koda:2011, Nishikawa:2019lsc, Turner:2021}.

Baryonic feedback~\citep{Bullock:2017xww,sales_baryonic_2022} and close encounters with nearby galaxies~\citep{moreno_galaxies_2022} provide additional mechanisms that can affect the evolution of halo centers within the CDM paradigm. However, the extent to which these processes can fully reproduce the diversity of observed galaxy properties remains an active area of research~\citep[see e.g.,][]{Santos_Santos_2020, Kaplinghat:2019dhn, Kong_2022}, and SIDM models that can explain the diversity provide well-motivated targets for investigation. 

The centers of dwarf galaxy halos are of particular interest as these systems are DM dominated and can undergo gravothermal collapse within the age of the Universe. Through gravothermal evolution, SIDM halos can span a range of central density profiles, from cores to collapsed cusps, with potential implications for interpreting small-scale observations~\citep{Kuzio_de_Naray_2014, Oman:2015xda, Relatores_2019, Santos_Santos_2020, Hayashi:2020jze, Fischer_2024, gutcke2025SIDM}. Consequently, dwarf galaxy halos with similar masses may exhibit substantially different central densities depending on their stage of gravothermal evolution~\citep{Kahlhoefer:2019oyt, Correa:2021, Roberts:2024uyw, fischer_dark_2026}.

To significantly impact dwarf galaxy central densities, the cross section per unit DM mass, \sm, must typically exceed $1$~\cmg~\citep{Spergel:1999mh, Dave:2001, Colin:2002, Vogelsberger:2012ku, Rocha:2012jg, Zavala:2012us, Vogelsberger2014, Fitts2019}. Matching observations of the most massive Milky Way~(MW) satellite galaxies further restricts to $\sigma/m > 5$~\cmg~\citep{Kaplinghat:2019svz, Correa:2020qam, Silverman_2022, Slone_2023, Zhang_2024}. %Valli:2017ktb
These considerations motivate studying large cross sections ($\sim10\text{--}100$~\cmg), which can drive dwarf galaxy halos beyond core formation and into the gravothermal-collapse regime~\citep{Turner:2021, Nadler_2021, Silverman_2022, Yang_2023_strong, fischer_dark_2026, engelhardt_marvelously_2026}.

The evolution of core-collapsing halos is described by the gravothermal fluid equations~\citep{lynden-bell_gravo-thermal_1968, balberg-SMBH-2002}. 
The fluid model has been shown to be consistent, within an order-unity scaling relation, with non-cosmological idealized N-body simulations~\citep{palubski2024numerical, Mace2024}. 
A parametric model based on the fluid equations has been shown to be consistent with N-body simulations of isolated MW-mass halos in the gravothermal-collapse regime~\citep{yang2024testing}.

Beyond the idealized setups, cosmological simulations provide a crucial tool for making SIDM predictions and testing analytic models~\citep{Vogelsberger2014, Robles_2017SIDMFIRE, Correa_2022, shen2022x, Fischer_2024, nadler2025, despali_aida-tng_2025, engelhardt_marvelously_2026}, however gravothermal collapse introduces numerical challenges that
must be addressed~\citep{palubski2024numerical, Mace2024, Fischer_2024_b}. 
As the core density rises, the local SIDM scattering timescale decreases, so a timestep chosen from the gravitational acceleration alone may be too large to resolve self interactions in the collapsing core. Accurately following this regime, therefore, requires a timestep criterion tied to the self-scattering probability. Recent convergence studies of idealized SIDM simulations show that default or CDM-motivated parameter choices can be insufficient for core collapse, where inadequate timesteps can stall or reverse the core evolution~\citep{palubski2024numerical, Mace2024, Fischer_2024_b}. These studies vary the DM mass resolution, timestep size, and gravitational softening length, providing convergence criteria that can be tested and adopted in cosmological SIDM simulations.

Implementing these findings, we present a suite of high-resolution cosmological zoom-in N-body simulations with refined time-stepping of halos that are predicted to undergo gravothermal collapse within the age of the Universe. We simulate six DM-only~(DMO) $\sim10^{10}$~\Msolar{} halos with initial conditions from the Feedback In Realistic Environments~(FIRE) project~\citep{Hopkins_2014,Hopkins_2018} using both CDM and SIDM with a constant cross section of $\sigma /m = 70$~\cmg. We analyze, for the first time, the density evolution of these dwarf-mass halos as they undergo gravothermal collapse in a cosmological setting and compare to  the fluid model. Our results demonstrate that merger history plays a key role in determining which halos collapse, introducing halo-to-halo variation that is not captured by the fluid model.

This paper is organized as follows. The simulation suite is introduced in Section~\ref{sec:suiteoverview}. Section~\ref{sec:merg} follows gravothermal collapse of the halos in a cosmological setting, focusing on the impact of mergers on the core density and on heat flow within the halo. Section~\ref{sec:simvsanalytic} compares halo cores and collapse timescales to the fluid model. We conclude in Section~\ref{sec:conclusion}. 
Appendix \ref{app:timestep} tests time-stepping parameters, 
Appendix \ref{app:res} discusses resolution limitations, 
Appendix \ref{app:corerad} compares the evolution of core radii with an isothermal profile, and 
Appendix \ref{app:diffSSIDM} presents the evolution of density and velocity dispersion profiles with two SIDM cross sections.

\section{Overview of Simulation Suite}
\label{sec:suiteoverview}

\begin{table*}[t]
\centering
\noindent\hspace*{-0.1\textwidth}\makebox[\textwidth]{%
\begin{tabular}{c l c c c c c}
    \hline
    \textbf{IC} & \textbf{Name} & $\bm{M_{\rm host}}$ & $\bm{\rho_{\rm s, CDM}}$ & $\bm{r_{\rm s, CDM}}$ & $\bm{t_{\rm \textbf{coll,mod}}}$ & $\bm{t_{\rm \textbf{end}}}$ \\
     &  & [$10^{10}$~\Msolar] & [$10^7~\rm M_\odot/kpc^3$] & [kpc] & [Gyr] & [Gyr] \\
    \hline
    
    \multirow{2}{*}{\centering m10q}
     & m10q-CDM & 1.01 & 2.53 & 2.72 & -    & 13.8 \\
     & m10q-s70 & 0.87 &  -    &   -   & 7.33 & 10.4 \\
    \hline

    \multirow{2}{*}{\centering m10d}
     & m10d-CDM & 1.09 & 3.67 & 1.78 & -    & 13.7 \\
     & m10d-s70 & 0.93 &  -    &   -   & 5.37 & 9.5 \\
    \hline
    
     \multirow{2}{*}{\centering m10i}
     & m10i-CDM & 1.77 & 6.79 & 1.66 & -    & 13.7 \\
     & m10i-s70 & 0.88 &  -    &   -   & 2.28 & 5.2 \\
    \hline
    
    \multirow{3}{*}{\centering m10b}
     & m10b-CDM & 1.23 & 2.02 & 2.31 & -    & 13.7 \\
     & m10b-s30  & 1.25 &  -    &  - & 22.30  & 13.7 \\
     & m10b-s70 & 1.26 &  -    &  -    & 10.12 & 9.0 \\
    \hline
    
    \multirow{2}{*}{\centering m10f}
     & m10f-CDM & 1.77 & 2.44 & 2.45 & -    & 13.7 \\
     & m10f-s70 & 1.32 &  -    &   -   & 7.18 & 6.4 \\
    \hline
    
    \multirow{2}{*}{\centering m10j}
     & m10j-CDM & 1.39 & 6.90 & 1.62 & -    & 13.7 \\
     & m10j-s70 & 0.84 &  -    &  -    & 2.28 & 4.9 \\
    \hline
    
    \end{tabular}
    }
    \caption{Summary of simulations generated in this work and their various parameters. From left to right, the columns refer to: 
    (1)~Halo initial condition~(IC);
    (2)~Simulation name, comprised of the IC and the DM model; 
    (3)~Bound mass within the virial radius ($r_{\rm 200m}$) of the host halo, $M_{\rm host}$, at the latest available timestep; 
    (4)~NFW scale density, $\rho_{\rm s, CDM}$, at $z=0$, shown only for CDM halos;
    (5)~NFW scale radius, $r_{\rm s, CDM}$, at $z=0$, shown only for CDM halos;
    (6)~Expected time at which the halo reaches gravothermal collapse, $t_{\rm coll, mod}=400\,t_{\rm c,0}$, as defined in Equation~\ref{eq:tco};
    and
    (7)~Time that the simulation is run to, $t_{\rm end}$. Redshift $z=0$ corresponds to $t_{\rm end}=13.8~(13.7)$~Gyr for simulations run in WMAP 7/9~(Planck 2014) cosmology.}
    \label{tab:simulation_runs}
\end{table*}

This work presents a suite of DMO zoom-in cosmological simulations of host halos of classical dwarf galaxies ($M_{\rm host} \sim 10^{10}$~\Msolar,\; hereafter ``m10s''), which are run with the \texttt{GIZMO} code using the  Tree+PM gravity solver~\citep{hopkins15, Hopkins_2018}. The suite samples a range of initial conditions~(ICs) to compare halo evolution in CDM and SIDM.
Table~\ref{tab:simulation_runs} summarizes the simulations and their key parameters. 

\subsection{SIDM Implementation}
\label{sec:sidm_imp}

DM self interactions are implemented in \texttt{GIZMO}~\citep{Fitts2019, Sameie_2021, Vargya_2022, Shen_2021, shen_dissipative_2024, Arora_2024} using the Monte Carlo-based scattering method from~\citet{Rocha:2012jg, Peter2013}, which is briefly summarized here. 
In this procedure, the DM simulation particle $i$ located at position $\textbf{r}$ is associated with a kernel $W(r, h_{s,i})$ with interaction smoothing length $h_{s, i}$.  The smoothing length roughly corresponds to the size of the particle and determines the number of nearest neighbors it can scatter with. We use the standard SIDM implementation in \texttt{GIZMO} where $h_{s, i}$ is adaptive to include 32 nearest neighbors and takes a minimum value of $0.3$~pc~\citep{Hopkins_2018, Shen_2021}.

The SIDM interaction rate between simulation particles $i$ and $j$ is 
\begin{equation}
    \Gamma_{ij} = (\sigma/m) \, m_{\rm \scriptscriptstyle DM} \, v_{\rm rel} \, g_{ij} \, ,
\end{equation} 
where \mdm~is the DM particle mass, $v_{\rm rel}$ is the relative velocity, and $g_{i j}$ is the estimated kernel overlap between the two particles. The latter is calculated using a cubic-spline kernel~\citep[e.g.,][]{Rocha:2012jg}.

The probability that particle $i$ scatters with any of its neighbors $j$ in the timestep $\delta t_i$ is 
\begin{equation}
P_{i j} = \Gamma_{i j} \, \delta t_i \, \leq \kappa \, ,
\label{eq:probij}
\end{equation}
where $\kappa$ is the maximum scattering probability at a given timestep. This parameter is introduced to limit the number of interactions per timestep to $\ll 1$.
The particle timestep, $\delta t_i$, is set adaptively; its maximum allowed value is
\begin{equation}
    \delta t_i = \sqrt{\frac{2 \,\eta \,h_{s,i}}{a}} \,,
    \label{eq:eta}
\end{equation}
and it is reduced when necessary to meet the criteria established in Equation~\ref{eq:probij}.
Here, $a$ is the magnitude of the particle's acceleration and $\eta$ controls the fraction of $h_{s,i}$ that the particle can travel after scattering in a given timestep. This parameter limits the displacement of particles with large scattering-induced accelerations between timesteps.

As shown in~\citet{palubski2024numerical, Mace2024}, SIDM halo evolution in these simulations is sensitive to two numerical parameters: $\eta$ and $\kappa$. We adopt $\eta=5 \times 10^{-4}$ and $\kappa=5 \times 10^{-4}$, both smaller than typical values in previous works~\citep{Robles_2017SIDMFIRE, Hopkins_2018, Correa_2022, palubski2024numerical, Mace2024, engelhardt_marvelously_2026}, to reduce numerical heating from multiple scattering events within a single timestep. Appendix~\ref{app:timestep} discusses the convergence tests that motivate these choices.

The SIDM simulations summarized in Table~\ref{tab:simulation_runs} have a constant cross section of $\sigma/m=70$~\cmg{} (hereafter ``s70''). An additional m10b run with $\sigma/m=30$~\cmg{} (hereafter ``s30'') is included to study how merger-induced core evolution depends on the SIDM cross section.
These cross sections are comparable to those predicted by viable velocity-dependent SIDM models at the characteristic relative velocities of m10 halos ($v_{\rm rel}\sim30\text{--}60$~\kms; \citealt{Correa:2021, Yang_2023_strong, Slone_2023, Nadler_2023}).
At the relevant velocity scales, s70 is consistent with MW-~\citep{Vogelsberger:2015gpr,Nadler:2020ulu} and cluster-scale constraints~\citep{2021JCAP...01..024S}, while being sufficient to drive a subset of dwarf galaxy halos and subhalos into the gravothermal-collapse regime~\citep[e.g.,][]{Lovell:2016nkp, Turner:2021}.

\subsection{Initial Conditions}
\label{sec:ICs}

The ICs sample diverse environments, with varied assembly histories, halo masses, and concentrations. We adopt halos from the m10 suite (m10q, m10d, m10i, m10b, m10f, m10j) drawn from the FIRE-2 project~\citep{Wetzel_2023, wetzel2025second} for continuity with prior CDM studies~\citep[e.g.,][]{Kim_2013,Hopkins_2014,Fitts:2016usl,Robles_2017SIDMFIRE,Hopkins_2018}. 
The ICs for halo m10q were first introduced by~\citet{Kim_2013}, while those for the remaining halos were first introduced by~\citet{Fitts:2016usl}. The m10q runs are evolved in a WMAP~7/9 cosmology~\citep{Komatsu_2011}, adopted from the AGORA collaboration~\citep{Kim_2013}: $\Omega_{\rm m} = 0.272$, $\Omega_\Lambda = 0.728$, $\Omega_{\rm b} = 0.0455$, and $H_0 = 70.2$~km~s$^{-1}$~Mpc$^{-1}$. 
The remaining halos use a slightly different cosmology, following the Planck Collaboration 2014 results~\citep{Planck2014}: $\Omega_{\rm m} = 0.266$, $\Omega_\Lambda = 0.734$, $\Omega_{\rm b} = 0.0449$, and $H_0 = 71$~km~s$^{-1}$~Mpc$^{-1}$. We adopt previously tested ICs rather than creating new ones with a consistent cosmology; effects of these cosmological differences are much smaller than halo-to-halo variation~\citep{Hopkins_2018}.

For the masses and concentrations considered, the fluid-model expectation is that each halo will undergo gravothermal collapse within a Hubble time with a constant cross section $\sigma/m=70$~\cmg. The collapse time for each halo estimated from the Navarro-Frenk-White~(NFW;~\citealt{Navarro:1996gj}) parameters of the corresponding CDM halo, $t_{\rm coll, mod}$, is provided in Table~\ref{tab:simulation_runs} and discussed further in Section~\ref{sec:fluidmodel}.  

While all CDM halos are evolved to $z=0$, this is not necessarily the case for their SIDM counterparts, as indicated by the end time, $t_{\rm end}$, in Table~\ref{tab:simulation_runs}. This is because, for an SIDM halo undergoing collapse, the central density can rise to values that dramatically increase computational cost. We terminate a run before $z=0$ if it does not reach the next snapshot within $1.1 \times 10^4$~CPU hours; the snapshot intervals range from $300$--$770$~Myr, depending on the termination time.
In all such cases, the host halo has either undergone gravothermal collapse or has dramatically deviated from the fluid model.
When the host halo has not collapsed, the increase in run time is due to a field halo within the simulation region undergoing gravothermal collapse.

The m10q halo is run with DM mass resolution \mdm{} of $1.5 \times 10^3$~\Msolar, while the other ICs are run with $m_{\rm \scriptscriptstyle DM}=3 \times 10^3$~\Msolar.\footnote{The variation in \mdm{} across ICs is due to simulation box size: m10q is run in a box with $5$~Mpc side length while the other simulations are run in $25$~Mpc boxes.
We adopt previously tested ICs to leverage existing validation and convergence studies and comment where resolution may impact results.} In all cases, gravity is softened using a cubic-spline kernel with a fixed Plummer-equivalent softening length $\epsilon_{\scriptscriptstyle\rm DM}=28$~pc, consistent with~\citet{Hopkins_2018,Wheeler_2019}. The choice of fixed rather than adaptive softening is to reduce numerical heating that can bias the evolution of both CDM and SIDM halos~\citep{Hopkins_2018,palubski2024numerical}. The value of $\epsilon_{\scriptscriptstyle\rm DM}$ meets the spatial- and mass-resolution criteria of~\citet{van_den_Bosch_2018} for resolving inner halo structure and the convergence criteria of~\citet{Mace2024} for gravothermal collapse in SIDM, minimizing numerical-relaxation effects.

\begin{table*}[t]
\centering
\noindent\hspace*{-0.1\textwidth}\makebox[\textwidth]{%
    \begin{tabular}{l c c c c}

    \hline
    \multicolumn{2}{c}{\textbf{Host Halos}} & \multicolumn{3}{c}{\textbf{Massive Radial Subhalos}} \\
    \textbf{Name} & \textbf{\# of mergers} & \bm{$\left(M_{\rm sub}/M_{\rm host}\right)_{\rm infall}$} & \bm{$\left(v_{\rm tan}/v_{\rm tot}\right)_{\rm infall}$} & \bm{$t_{\rm infall}$} \\
    
    & ($M_{\rm sub}/M_{\rm host} \geq 0.01$) & & & \\
    \hline
    \multirow{2}{*}{m10q-s70} & \multirow{2}{*}{18}
    & 0.17 & 0.174 & 2.0 \\
    & & 0.10 & 0.079 & 1.2 \\
    \hline
    m10d-s70 & 18 & - & - & - \\
    \hline
    m10i-s70 & 6 & - & - & - \\
    \hline
    \multirow{4}{*}{m10b-s70} & \multirow{4}{*}{15}
    & 0.37 & 0.059 & 7.7 \\
    & & 0.22 & 0.187 & 3.3 \\
    & & 0.11 & 0.099 & 5.7 \\
    & & 0.10 & 0.098 & 1.6 \\
    \hline
    m10f-s70 & 17 & 0.16 & 0.136 & 3.6 \\
    \hline
    m10j-s70 & 12 & - & - & - \\
    
    \hline
    \end{tabular}
    }
    \caption{Merger properties in each SIDM host halo. The first two columns correspond to the host halo and the last three correspond to subhalos that have $M_{\rm sub}/M_{\rm host} \geq 0.1$ and $v_{\rm tan}/v_{\rm tot} \leq 0.2$. Columns show: (1)~Host name; (2)~Total number of mergers with infall mass of at least $0.01 \, M_{\rm host}$; (3)~Subhalo-to-host halo mass ratio at subhalo infall time; (4)~Tangential velocity, normalized by total velocity, at infall; and (5)~Infall time.
    }
    \label{tab:mergers}
\end{table*}

\cite{Wheeler_2019} find that CDM cosmological hydrodynamic simulations of $10^{9-10}$~\Msolar{} halos are converged at the DM particle-mass resolutions used here. Unfortunately, resolution convergence for the SIDM halos in our suite is harder to verify. \cite{palubski2024numerical, Mace2024} find that in idealized SIDM simulations, halos with $\geq 5\times 10^5$ particles reproduce results from the fluid model and minimize numerical scatter in core density and gravothermal-collapse timescales. As shown in these works, running at too-low resolution can artificially decrease the minimum core density and increase the gravothermal-collapse time. All host halos in our suite match the particle-number criterion derived from isolated simulations, with over $10^6$ bound particles within their virial radii ($r_{\rm 200m}$) by $t_{\rm end}$ and at least $5\times 10^5$ particles within $r_{\rm 200m}$ after the first $\sim 2$~Gyr. It is unclear, however, whether and how this criterion generalizes to cosmological halos.
To assess this, Appendix~\ref{app:res} demonstrates how the density profiles of m10i-s70 and m10b-s70 are altered by running up to two steps lower in resolution. By examining both a halo that collapses with minimal impact from mergers (m10i) and one that does not collapse due to mergers (m10b), we find evidence that the impact of mergers delaying collapse and decreasing core densities is physical and largely resolution-independent while the normalization of the density and timing of the gravothermal evolution may shift with improvements to resolution.

\subsection{Host and Subhalo Properties}
\label{sec:properties}

A key focus of this work is to examine the evolution of the host halo's density profile in response to subhalo mergers. To this end, we track the time evolution of the core density, $\rho_c$, of each host by fitting its inner density profile to the analytic expression~\citep{burkert1995structure}
\begin{equation}
    \rho(r) = \rho_{\rm c} \left(1 + \left(\frac{r}{r_0} \right)^2\right)^{-3/2}  \, ,
    \label{eq:coreden}
\end{equation}
where $r_0$ is the scale radius where the slope of the log-profile becomes $-3$. We define $r = 0$ as the core's center of mass, calculated using particles within $r_{\rm 200m}$ that possess the highest local DM density. 
To remain independent of resolution, following~\citet{palubski2024numerical}, the core's center of mass is calculated using 0.67\% of particles within $r_{\rm 200m}$.
To ensure dense subhalos do not bias this calculation, we also determine the core density using the merger-tree-identified center and adopt whichever location yields the higher central density. This specific centering procedure is critical since the core can slosh around within the host~\citep{palubski2024numerical}. 
% Using the center of mass of the entire halo can produce erroneous results, as seen in the m10q-s70 simulation, where it underestimates the core density by an average factor of $0.73$ over time.

We identify DM subhalos in the simulations using the \texttt{Rockstar} halo finder~\citep{behroozi13a}. The merger trees linking these subhalos across the simulation snapshots are constructed using the \texttt{consistent-trees} algorithm~\citep{behroozi13b}, following the method described by~\cite{Samuel_2019, samuel2021planes}. The merger trees can reliably track subhalos with masses~$\geq~10^{5}$~\Msolar{} at the fiducial resolution. However, subhalo pericenters are not reliably tracked at this snapshot cadence; we therefore classify subhalos by their parameters at infall, when the subhalo first enters $r_{\rm 200m}$ of the host. Throughout this work, mergers are defined as halos that have passed within $r_{\rm 200m}$ of the host halo.

We expect massive subhalos on radial orbits to have the largest impact on the host's core evolution as they are more likely to reach its vicinity. In doing so, these mergers introduce a new heat flow source beyond SIDM scatterings. We summarize the merger history of the subhalos in Table~\ref{tab:mergers}.\footnote{We do not show halo m10b-s30 because its merger history is nearly identical to m10b-s70, differing only by three additional mergers, all with $M_{\rm sub}/M_{\rm host} \leq 0.03$. For comparison, m10b-s70 experiences seven mergers in this mass-ratio range.} For each host, we list the total number of mergers with $M_{\rm sub}/M_{\rm host} \geq 0.01$, where $M_{\rm sub}$ is the bound mass within $r_{\rm 200m}$ of the subhalo at infall and $M_{\rm host}$ is the bound mass within $r_{\rm 200m}$ of the host at the time of subhalo infall. 
We then identify the subset of mergers expected to be most relevant for core evolution: massive radial mergers with $M_{\rm sub}/M_{\rm host} \geq 0.1$ and $v_{\rm tan}/v_{\rm tot} \leq 0.2$, where $v_{\rm tan}$ is the tangential velocity of the subhalo at infall and $v_{\rm tot}$ is its total velocity at infall. We note that this mass threshold also reflects the resolution limit of the halo finder, below which subhalo identification becomes incomplete and may preferentially miss low-mass mergers in late-forming halos. Since the snapshot cadence is too coarse to resolve pericenters, we use $v_{\rm tan}/v_{\rm tot}$ as a proxy for orbital radiality. 

While $v_{\rm tan}/v_{\rm tot}$ provides an estimate of how radial the orbit would be in a static spherical potential at infall, it need not be a good determinant of the pericenter distance, which is what we need to track for energy input into the core. The snapshots we have saved do not have the time resolution to compute the pericenter distance directly for all subhalos. Factors including non-spherical mass distributions, dynamical friction (relevant for large-mass-ratio mergers), drag created by the interactions of DM particles in the subhalo with the main halo, and evolving potential make $v_{\rm tan}/v_{\rm tot}$ a rough proxy for the pericenter distance.

We define radial mergers as those below the median value of $v_{\rm tan}/v_{\rm tot}$ among mergers with $M_{\rm sub}/M_{\rm host} \geq 0.01$, which corresponds to $v_{\rm tan}/v_{\rm tot} \leq 0.2$. 
As will be demonstrated, the formation history of each host plays an integral role in determining whether a halo undergoes gravothermal collapse.

\section{Gravothermal Collapse in a Cosmological Setting}
\label{sec:merg}

\begin{figure*}[t]
    \centering
    \includegraphics[width=\textwidth]{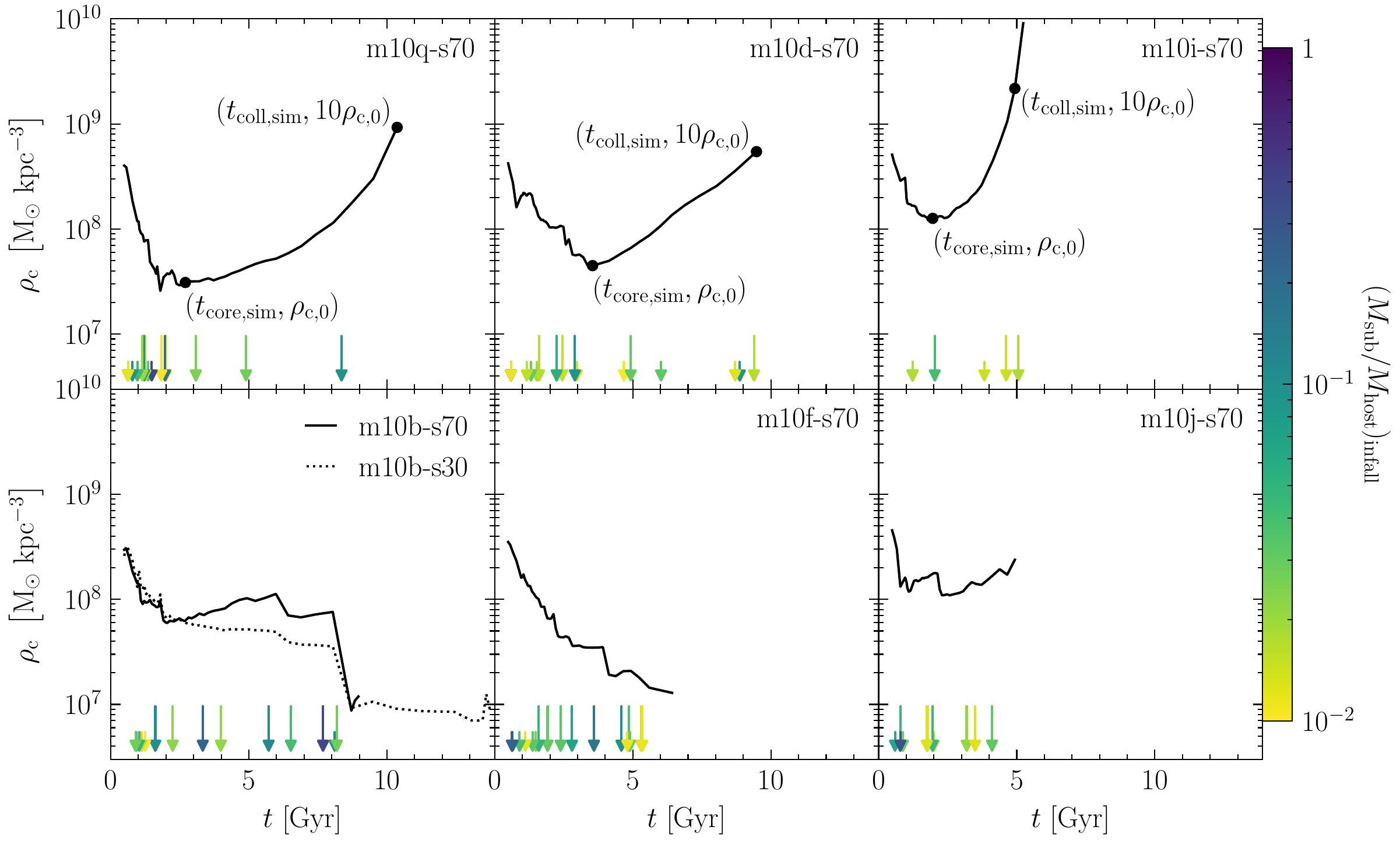}
    \caption{Time evolution of the central halo's core density. Arrows along the horizontal axis mark subhalo infall times and are color coded by the subhalo-to-host mass ratio at infall. Their lengths indicate orbital radiality, with long arrows corresponding to radial mergers ($v_{\rm tan}/v_{\rm tot} \leq 0.2$). Halos that undergo gravothermal collapse are shown in the top row, while those that do not are in the bottom row; the bottom-left panel additionally compares the evolution of m10b-s70 to m10b-s30. For collapsing halos, we mark the minimum core density at $(t_{\rm core, sim}, \rho_{\rm c,0})$ and define collapse at $(t_{\rm coll, sim}, 10\,\rho_{\rm c,0})$.}
    \label{fig:coreden}
\end{figure*}

This section explores the density evolution of the SIDM halos in our suite. Section~\ref{sec:core_mergers}  presents the core evolution of each host, and Section~\ref{sec:mass_energy} discusses how the results correlate with the flow of heat through the halo.

\subsection{The Impact of Mergers on Core Density}
\label{sec:core_mergers}

Figure~\ref{fig:coreden} plots the time evolution $\rho_c$ for the s70 runs. The top row shows the evolution of halos m10q,d,i, which undergo gravothermal collapse within a Hubble time, as predicted.  The bottom row shows the evolution of halos m10b,f,j, which do not undergo gravothermal collapse despite being expected to do so (see Section~\ref{sec:ICs}). The arrows mark the times of merger events; their color encodes the merger mass ratio, while their length indicates the orbital radiality (long arrows indicate radial mergers, which have $v_{\rm tan}/v_{\rm tot} \leq 0.2$).

Halos m10q,d,i first undergo core expansion until the density reaches a minimum, $\rho_{\rm c, 0}$.\footnote{To account for the noise in the data at early times (e.g., the dips in m10q-s70 at $1\text{--}2$~Gyr), we smooth the core density evolution using a moving average with window size of 0.6 Gyr to calculate $t_{\rm core, sim}$ and $\rho_{\rm c, 0}$. These smoothed values are used throughout this work while figures show the unsmoothed evolution for reference.} This occurs at $t_{\rm core, sim}\simeq 3,4,2$~Gyr, respectively. 
After this point, the density rises rapidly as each halo enters the gravothermal-collapse regime. Because the simulations cannot track the collapse process to its ultimate end (e.g., $\rho_{\rm c} \rightarrow \infty$), we define a proxy for the collapse time, $t_{\rm coll, sim}$, to be the earliest snapshot for which $\rho_{\rm c}\geq 10\rho_{\rm c,0}$. 
Future simulations with more efficient and accurate treatments of self interactions at high densities~\citep[e.g.,][]{fischer_cosmological_2026} are needed to follow collapse beyond this point.
Halos m10q,d collapse by $t_{\rm coll, sim}\simeq 10$~Gyr, while m10i collapses in roughly half the time.\footnote{For both m10q and m10d, $t_{\rm coll, sim}$ occurs at the latest simulation snapshot, although the density at this time is already well above $10\,\rho_{\rm c, 0}$ This is because the density at the penultimate snapshot has not yet reached $10\,\rho_{\rm c, 0}$.} 

As seen in Table~\ref{tab:mergers}, m10i has only six mergers with $M_{\rm sub}/M_{\rm host} \geq 0.01$, while all other halos in the suite have more than 12 mergers in this mass range. Halo m10i also stands out for having no mergers with mass ratio $\geq 0.1$. Because it has the most quiescent history, we expect it to closely follow the fluid-model prediction for gravothermal evolution---which is indeed the case, as will be demonstrated in Section~\ref{sec:fluidmodel}. 
Halos m10q,d each have 18 mergers with $M_{\rm sub}/M_{\rm host} \geq 0.01$---the largest number in the suite---yet these mergers do not prevent the halos from collapsing (see Figure~\ref{fig:coreden}), likely due to their low masses, circular orbits, and early infall times. Halo m10d has no massive radial mergers to impact its core density and m10q has two massive radial mergers, which lead $\rho_{\rm c}$ to decrease by a factor of $\sim 1.3$--$1.6$ after each merger, but do not prevent the halo from undergoing gravothermal collapse, as discussed later.

Halos m10b,f,j do not collapse, despite being expected to do so. The result for m10b-s70 is particularly striking, as it experiences a period from $\sim 2\text{--}6$~Gyr when the core density increases, before a significant $1:3$ merger at $7.7$~Gyr (together with other, less-massive mergers) causes a sharp decrease in $\rho_{\rm c}$. Given this dramatic decrease in $\rho_{\rm c}$, m10b provides a useful case study to investigate how different SIDM cross sections impact $\rho_{\rm c}$ evolution. 
Comparatively, m10b-s30 does not experience a period of increasing core density (a slower gravothermal evolution is expected at lower cross sections) but does experience a $1:3$ merger at $7.7$~Gyr, which causes a sharp decrease in $\rho_{\rm c}$---in this case, by a factor of $3.9$ compared to $8.7$ in m10b-s70. 
In the m10b-s30 run, the decrease in $\rho_{\rm c}$ is sustained until $z=0$.
 
In contrast, m10f exhibits a more steady decline in its core density, likely because it experiences a steady barrage of closely-spaced $\sim1:30$ mergers as well as one massive radial merger. While m10j does not reach $10 \, \rho_{\rm c, 0}$ by $2.28$~Gyr as expected, it does show signs of starting to collapse, indicating that it is early in the gravothermal-collapse regime (see Section \ref{sec:mass_energy}).
We do not run the simulation past 4.9~Gyr due to increased run time from other halos in the cosmological box developing high-density cores.  

\begin{figure*}
    \centering
    \includegraphics[width=\textwidth]{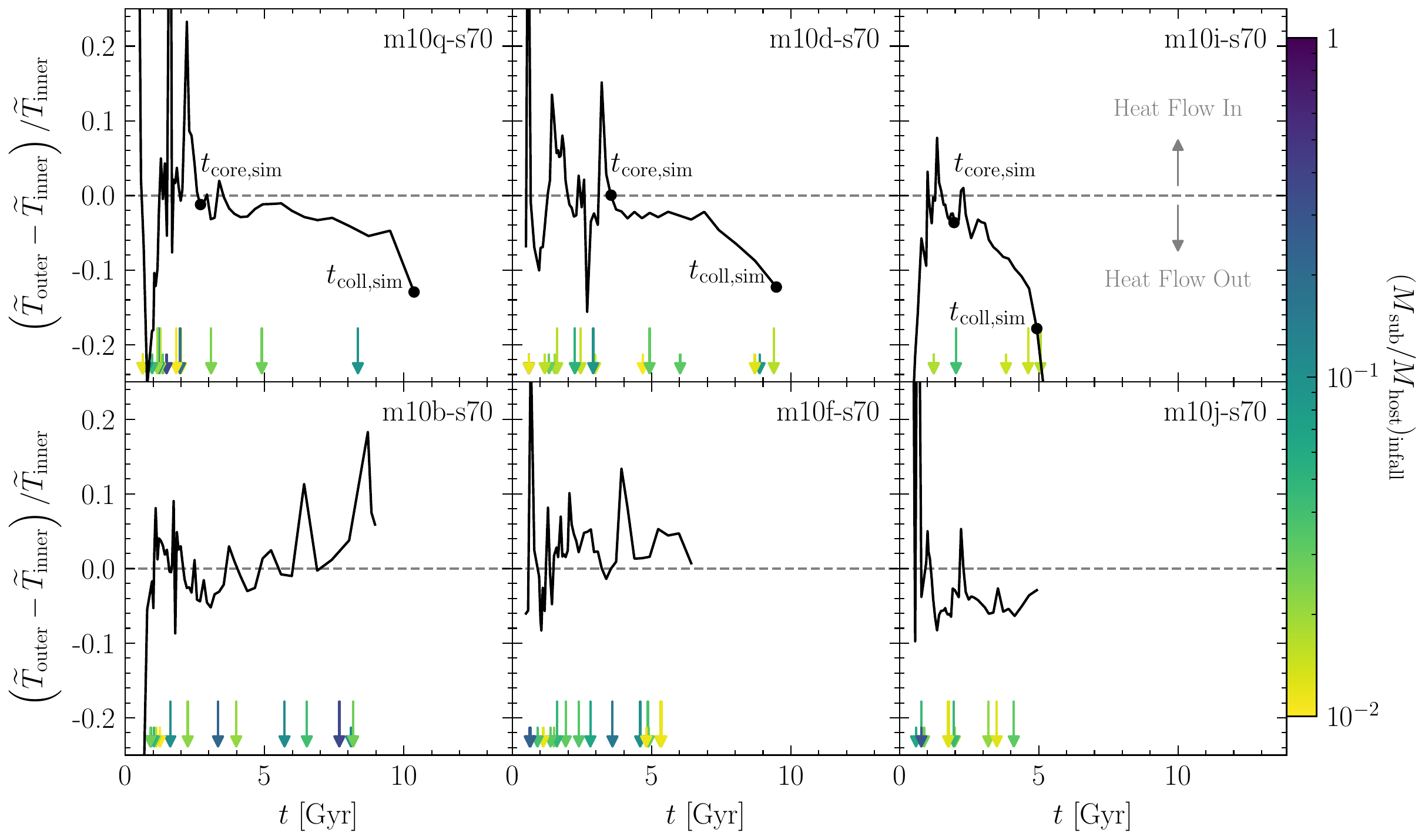}
    \caption{Fractional change in average temperature between inner and outer regions as a function of time. Arrows along the horizontal axis mark merger infall times and are colored by the subhalo-to-host mass ratio at infall. Their lengths indicate orbital radiality, with longer arrows corresponding to more radial mergers. The average temperature $\widetilde{T}$ is defined in Equation~\ref{eq:avgt}, with $\widetilde{T}_{\rm inner}$ and $\widetilde{T}_{\rm outer}$ computed within 1~kpc and 1--5~kpc of the halo center, respectively. Positive (negative) fractional differences correspond to $\widetilde{T}_{\rm inner} > \widetilde{T}_{\rm outer}$ ($\widetilde{T}_{\rm inner} < \widetilde{T}_{\rm outer}$). Halos are arranged with collapsing systems in the top row and non-collapsing systems in the bottom row. As in Figure~\ref{fig:coreden}, the time of minimum core density ($t_{\rm core, sim}$) and core collapse ($t_{\rm coll, sim}$) are marked for collapsing halos.} 
    
    \label{fig:coreenergy}
\end{figure*}

\subsection{The Impact of Mergers on Heat Flow}
\label{sec:mass_energy}

A characteristic signature of a core-collapsing halo is the temperature gradient of a kinematically hot core relative to a colder outer halo. For mergers to disrupt the gravothermal evolution, they must raise the temperature of the outer halo above the core temperature. The halo will then be unable to core collapse until it is given time to reach thermal equilibrium without further temperature changes from other mergers. As shown in Section~\ref{sec:core_mergers}, this disruption can occur through different merger histories. For m10b, a single $1:3$ merger is sufficient to significantly delay collapse, while for m10f, a higher frequency of lower-impact mergers produces a similar effect. In both cases, the underlying physical effect is a merger-driven change in the halo temperature structure.

To quantitatively measure these temperature changes, %To quantify the direct impact of mergers on heat flow, 
we calculate the average temperature (specific kinetic energy) of DM particles within region $S$, normalized by the mass within $S$, defined as 
\begin{equation}
    \widetilde{T} = \frac{1}{M}\frac{E_{\rm kin}}{m_{\scriptscriptstyle \rm DM}} = \frac{1}{2M}\sum_{i=1}^N v_i^2 \, ,
    \label{eq:avgt}
\end{equation}
where $E_{\rm kin}$ is the total kinetic energy, $v_i$ is the velocity of the $i^{\rm th}$ particle, and $N$ and $M$ are the number of particles and total mass within $S$, respectively. We compare the average temperature within $1$~kpc, $\widetilde{T}_{\rm inner}$, with that measured over the radial range $1$--$5$~kpc, $\widetilde{T}_{\rm outer}$. The inner radius is chosen to approximately match the characteristic core size, which is typically $\lesssim 1$~kpc (see Figure \ref{fig:corerad} for details).
The outer boundary at $5$~kpc encompasses twice the largest core radius reached by any of the six halos. Varying the outer radius changes the magnitude of the inferred heat flow but not its direction; since our analysis focuses on the latter, this choice does not affect our conclusions.

Figure~\ref{fig:coreenergy} shows the evolution of $(\widetilde{T}_{\rm outer} - \widetilde{T}_{\rm inner})/\widetilde{T}_{\rm inner}$ for the six halos, where a negative (positive) value indicates heat flowing outward from (inward towards) the core. Inward heat flow heats the core, causing it to expand and delaying the onset of gravothermal collapse. Outward heat flow drives gravothermal collapse. As in Figure~\ref{fig:coreden}, the arrows mark merger events.

The three collapsing halos (top row) all develop persistently negative temperature gradients after an initial period of turbulence associated with early halo formation. The late-time evolution of m10q,d is relatively quiet, with mergers after $\sim 3$~Gyr having little effect on the direction of heat flow. Halo m10i, which experiences the fewest mergers in the suite, sustains the strongest outward heat flow and collapses more rapidly.

The three non-collapsing halos (bottom row) are distinguished by recurring episodes of inward heat flow, each correlated with subhalo infall. Halo m10b provides the clearest example: the direction of heat flow reverses several times, with several pronounced inward spikes, the largest of which coincides with the major merger at $7.7$~Gyr. Appendix~\ref{app:diffSSIDM} compares the evolution of the velocity-dispersion profiles of m10b for $\sigma/m=30$ and $70$~\cmg{} to illustrate how the impact of the major merger depends on SIDM cross section. Halo m10f likewise sustains considerable inward heat flow throughout its evolution from frequent mergers. Halo m10j exhibits outward heat flow at the latest snapshots, indicating that it is early in the gravothermal-collapse regime. 

\subsection{Diagnostics for Collapse}
\label{sec:diagnostics}

The presented SIDM suite suggests several diagnostics for predicting whether a halo will undergo collapse based on its merger history: subhalo-to-host mass ratio at infall, merger radiality at infall, maximum quiescence time interval, and time of merger.  Each of these is now discussed in turn.

First, consider the mass ratio and radiality of a merger.  As indicated in Table~\ref{tab:mergers}, the halos associated with the most massive ($M_{\rm sub}/ M_{\rm host} \geq 0.1$) and most radial ($v_{\rm tan}/v_{\rm tot} \leq 0.2$) mergers are m10q,b,f. 
Neither m10b nor m10f collapse.  The former undergoes four mergers within this range, including a major $1:3$ event---the most significant experienced by any halo in the suite. The latter contains one subhalo in this regime and is also notable for having the highest frequency of mergers with $v_{\rm tan}/v_{\rm tot} \lesssim 0.2$ across the entire suite (as seen from the arrows in Figure~\ref{fig:coreden}). The fact that m10q does collapse despite having several massive, radial mergers suggests other variables at play---likely merger frequency/timing.

Notably, radial mergers that have $M_{\rm sub}/ M_{\rm host} \leq 0.1$ do not significantly disrupt the $\rho_{\rm c}$ evolution. For example, the three radial mergers experienced by m10i, which all have $M_{\rm sub}/M_{\rm host} < 0.04$, do not cause a decrease in $\rho_{\rm c}$. This indicates that mass may be more important than orbital radiality in determining merger impact.

Second, consider how the merger frequency affects the gravothermal evolution.  As a proxy for this, we quantify the maximum quiescence time interval between the mergers, normalized by the collisional relaxation timescale, $\Delta t_{\rm max}/t_{\rm relax}$. The relaxation timescale is defined as~\citep{Balberg:2002ue, Outmezguine_2023, Yang_2022}
\begin{equation}
t_{\rm relax}^{-1}= \sqrt{64G} \, \frac{\sigma}{m} \, \rho_{\rm s, CDM}^{3/2} \, r_{\rm s, CDM} \,  , 
\end{equation}
where $G$ is the gravitational constant, $\rho_{\rm s, CDM}$ is the NFW scale density, and $r_{\rm s, CDM}$ is the scale radius. 
We find that m10b,f have  $\Delta t_{\rm max}/t_{\rm relax} < 100$, while m10q,d,i,j have $\Delta t_{\rm max}/t_{\rm relax}>250$. Because it takes $100\text{--}150\,t_{\rm relax}$ for an idealized, isolated halo to reach the end of the core-expansion phase~\citep{Essig:2018pzq,Nishikawa:2019lsc,Outmezguine_2023}, halos with frequent mergers on timescales less than that may not collapse. 

Lastly, consider the importance of merger timing. Mergers occurring before $t_{\rm core, sim}$ may not prevent gravothermal collapse if the energy they inject is aligned with the inwards heat flow that is characteristic of the core-expansion phase. Halo m10q presents an excellent example: despite experiencing two mergers in the massive and radial regime, it still undergoes gravothermal collapse. 
These mergers occur at $0.4$ and $0.7\,t_{\rm core, sim}$. Halo m10d also experiences a radial merger with $M_{\rm sub}/M_{\rm host} = 0.09$ at $0.7\,t_{\rm core, sim}$, which is followed by collapse.  Figure~\ref{fig:coreenergy} explicitly shows that these specific mergers align with the inwards heat flow for these halos.
%A similar trend is shown in m10b. When $t_{\rm core, sim}$ is defined as the local minimum of $\rho_{\rm c}$ before its major merger at $7.7$~Gyr, its earliest massive radial mergers occur at $0.8$ and $1.7\,t_{\rm core, sim}$. These mergers do not prevent the increase in $\rho_{\rm c}$ from $\sim2$--$6$~Gyr. Mergers before about $t_{\rm core, sim}$ do not prevent gravothermal collapse, likely because the energy they inject into the halo is aligned with the inwards heat flow that is characteristic of the core-expansion phase.

Halos can also experience mergers well after the onset of core collapse without disrupting their evolution. Halo m10q experiences a merger with $M_{\rm sub}/M_{\rm host} \sim 0.1$ at $0.8 \ t_{\rm coll, sim}$. Similarly, m10d experiences three mergers after $0.9 \ t_{\rm coll, sim}$, a massive one with $M_{\rm sub}/M_{\rm host} \sim 0.1$ and two with mass ratio $M_{\rm sub}/M_{\rm host} \sim 0.01$. Lastly, m10i experiences two massive mergers after $0.6  \ t_{\rm coll, sim} $ with $M_{\rm sub}/M_{\rm host} \sim 0.1$. Furthermore, all the mergers stated are either classified as radial or have $v_{\rm tan}/v_{\rm tot} < 0.22$. These halos collapse despite these seemingly significant mergers. This could be due to the fact that once a significant temperature gradient is established, it takes more injected energy to disrupt it. However, given our sample size, we cannot rule out the possibility of halos deep in the core-collapse regime reverting back to the core-expansion phase.

The results presented in this section suggest that the subhalo-to-host mass ratio, merger orbital radiality, merger frequency, and specific timing of mergers play a critical role in determining the halo's gravothermal evolution and motivate follow-up studies using non-cosmological simulations to systematically vary over a merger's properties, recording the impact on the host's gravothermal evolution.

\section{Comparisons to Fluid Model}
\label{sec:simvsanalytic}

Having explored the role of mergers in modifying the density evolution, we proceed to compare the simulated core evolutions with the fluid-model predictions.

\subsection{Fluid Model}
\label{sec:fluidmodel}

The fluid model evolves a spherically symmetric, isolated, and virialized halo through time by solving mass continuity, hydrostatic equilibrium, and heat flux equations~\citep{lynden-bell_gravo-thermal_1968, balberg-SMBH-2002, lynden-bell_1980, Essig:2018pzq}. The model agrees, up to an $\mathcal{O}(1)$ factor, with semi-analytic and idealized N-body simulations of isolated~\citep{Koda:2011, Jiang_2023, Yang_2023_Gravothermal} and tidally stripped halos~\citep{Nishikawa:2019lsc, zeng2021corecollapse}.
\cite{Outmezguine_2023} use this model to describe the gravothermal collapse process in isolated SIDM halos within the long mean-free path regime, and \cite{gadnasr2023} extend it into the short mean-free path regime.\footnote{The long mean-free path regime is defined such that the SIDM particle mean-free path is greater than the halo gravitational scale height, $H=v/\sqrt{4 \pi G \rho}$, where $v$ is the one-dimensional velocity dispersion and $\rho$ is the halo density.} The halos in this work begin and remain in the long mean-free path regime.

In the long mean-free path regime, the solution is self-similar. Thus, the time it takes a halo to undergo gravothermal collapse is scale invariant~\citep{Nishikawa:2019lsc, Outmezguine_2023, Yang_2024_parametric}. The relevant time scale is the collisional timescale associated with the interactions~\citep{Outmezguine_2023}
\begin{equation}
\label{eq:tco}
    t_{\rm c,0} = \frac{2}{3 \, a \, C} \left( \rho_{\rm c,0} \, v_{\rm c,0} \, \frac{\sigma}{m}\right)^{-1}.
\end{equation}
Here, $\rho_{\rm c, 0}$ and $v_{\rm c, 0}$ are the core density and velocity dispersion at the time of minimum core density, $a = 4/\sqrt{\pi}$, and $C$ is an order-unity constant that depends on the heat conductivity. While $C$ is typically determined by calibrating to simulations, we do not fit for it in this work because, as discussed in the next subsection, the s70 halos do not follow the analytic model well enough. Previous studies using idealized N-body simulations find $C=0.6$--$0.84$~\citep{Essig:2018pzq, Nishikawa:2019lsc, palubski2024numerical, Yang_2024_parametric}; we adopt the geometric mean of this range, $C=0.71$. Using $C=0.6$~($0.84$) would increase~(decrease) predicted timescales by $15\%$.

To increase the predictive power of the fluid model,~\citet{Outmezguine_2023} show that $\rho_{\rm c, 0}$, $v_{\rm c, 0}$, and the core radius at the time of minimum core density, $r_{\rm core ,0}$, can be derived directly from the initial NFW parameters. Specifically, they find $\rho_{\rm c,0}/\rho_{\rm s, CDM}=2.4$, $r_{\rm core ,0}/r_{\rm s, CDM} = 0.45$, and $v_{c,0}/V_{\rm max, CDM} = 0.64 $ where $V_{\rm max, CDM}$ is the maximum circular velocity of the corresponding NFW profile. Following convention, we define the time to reach minimum core density as $t_{\rm core, mod} = 50\,t_{\rm c,0}$~\citep{Outmezguine_2023},\footnote{Values in the range 50--70~\citep{Outmezguine_2023, palubski2024numerical} have been adopted in the literature, implying a possible increase of up to $40\%$ in $t_{\rm core, mod}$ relative to our fiducial value.}
and the time to reach gravothermal collapse as $t_{\rm coll, mod} = 400\, t_{\rm c,0}$~\citep{Outmezguine_2023, palubski2024numerical}.\footnote{Values in the range $300$--$400$ have been adopted in the literature~\citep{balberg-SMBH-2002,Nishikawa:2019lsc,Gurian_2025,palubski2024numerical,Outmezguine_2023}, 
implying a possible decrease of up to $25\%$ in $t_{\rm coll, mod}$ relative to our fiducial value.
}

The fluid model will be most accurate when the system is in hydrostatic equilibrium and is thermalizing. This is most likely to be violated at early times when the halo happens to be in the core-expansion phase simply because mergers with larger mass ratios are expected then. For this reason, we divide the comparison between the fluid model and the simulations into two regimes: the core-expansion evolution leading to minimum core density and the subsequent evolution toward gravothermal collapse.

\subsection{Comparing to Simulations}
\label{sec:simvsmod}

\begin{figure*}
    \centering
    \includegraphics[width=\linewidth]{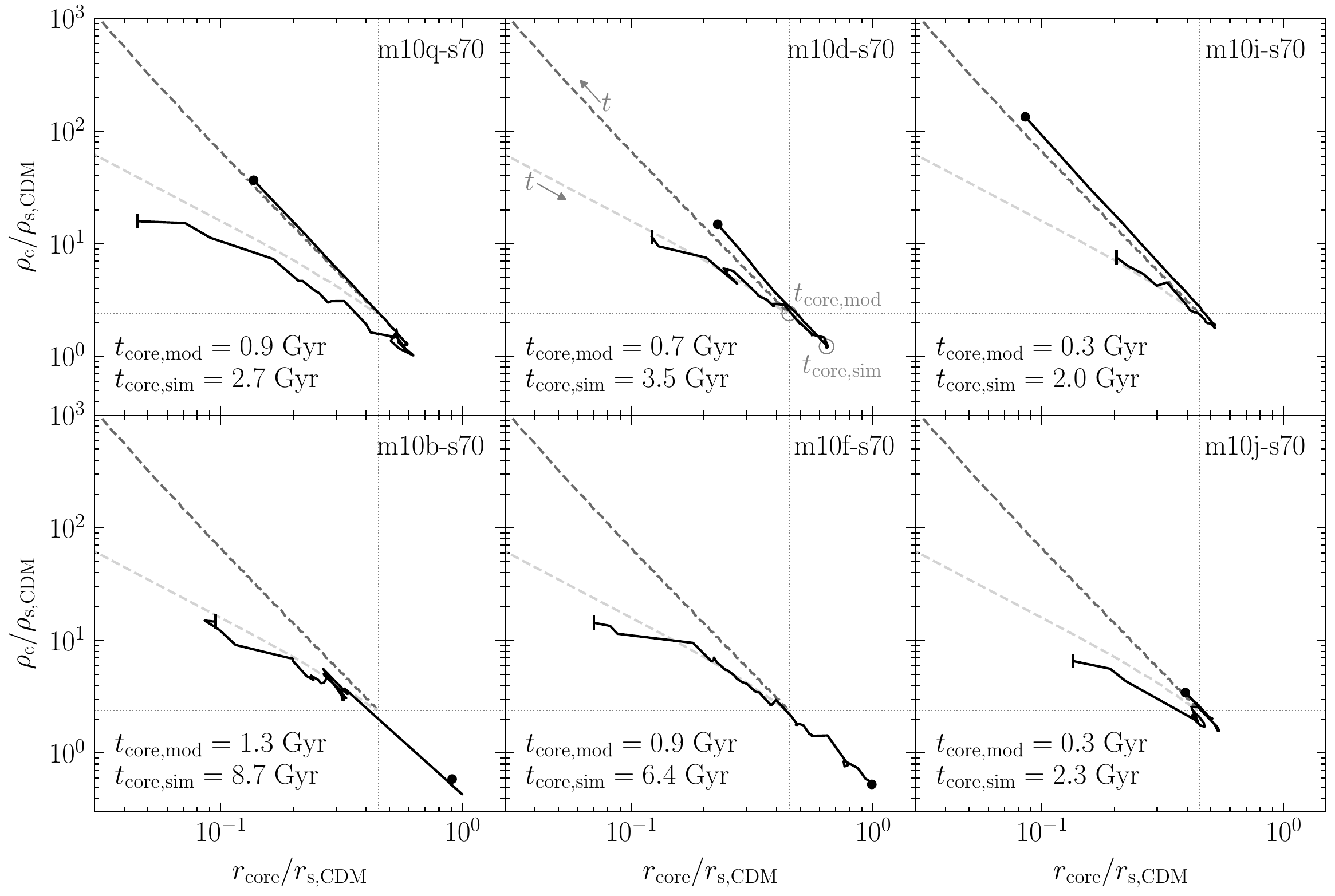}

    \caption{Evolution of core density and core radius for simulated halos compared to the fluid model. The dashed lines are the $\rho_c/\rho_{s,\rm{CDM}}$ and $r_{\rm{core}}/r_{\rm{s,CDM}}$ evolutions as predicted by the fluid model in~\citealt{Outmezguine_2023}, separated by the core-expansion phase (light gray) and core-collapse phase (dark gray). The core density is normalized by the CDM scale density at $z=0$, $\rho_{\rm s,CDM}$, and the core radius by the CDM scale radius, $r_{\rm s,CDM}$. The earliest timestep of each halo is marked by a vertical line, and the latest timestep by a dot. The direction of evolution is indicated by gray arrows along the model curve in the top middle panel. Dotted gray lines mark the normalized values $\rho_{\rm c,0}/\rho_{s,\rm{CDM}} = 2.4$ and $r_{\rm core,0}/r_{\rm{s,CDM}} = 0.45$ in the fluid model. The predicted time to reach this point, $t_{\rm core,mod}$, is shown in the bottom left corner of each panel, along with the corresponding simulated time, $t_{\rm core,sim}$.
    }
    \label{fig:analyticrhocrc}
\end{figure*}

This subsection compares the fluid model to simulations using the \texttt{GravothermalSIDM}\footnote{\url{https://github.com/kboddy/GravothermalSIDM}} code~\citep{Outmezguine_2023, gadnasr2023} to analytically evolve an SIDM halo from an initial NFW profile through gravothermal collapse. 
We use the m10q-CDM density profile at $z=0$ as the initial condition for the fluid-model realization and normalize by $\rho_{\rm s, CDM}$ and $r_{\rm s, CDM}$ so that the result is scale invariant. This semi-analytic prediction is then compared to the simulated evolution of each halo in the suite, normalized by its corresponding $\rho_{\rm s, CDM}$ and $r_{\rm s, CDM}$. We obtain the scale density, scale radius, and maximum circular velocity for a CDM halo by fitting an NFW profile to its density distribution between $r=0.5$--$30$~kpc using \texttt{COLOSSUS}~\citep{Diemer_2018}. The $\rho_{\rm s, CDM}$ and $r_{\rm s, CDM}$ values of these profiles are provided in Table~\ref{tab:simulation_runs}.

Figure~\ref{fig:analyticrhocrc} compares the $\rho_{\rm c}/\rho_{\rm s, CDM}$ and $r_{\rm core}/r_{\rm s, CDM}$ values predicted by the fluid model with the results from our simulations. 
For the core-expansion phase, the simulations do not consistently follow the fluid-model predictions at or prior to $\rho_{\rm c,0}$. The three halos that undergo collapse (top row) develop cores with larger radii and lower densities than predicted; among them, m10i shows the closest agreement, reaching $\rho_{\rm c,0}/\rho_{\rm s, CDM}=1.8$ and $r_{\rm core,0}/r_{\rm s, CDM}=0.5$. Its central density is therefore lower by only a factor of 1.3, while its core radius is larger by only a factor of 1.1, relative to the fluid-model prediction.
Comparatively, m10q,d reach $\rho_{\rm c,0}/\rho_{\rm s, CDM}=1.3$, $1.2$ and $r_{\rm core,0}/r_{\rm s, CDM}=0.6$, $0.6$, respectively. Their central densities are therefore lower by a factor of 1.9 and 2.0, while their cores are larger by a factor of 1.3 and 1.4, relative to the fluid-model expectation.

For the non-collapsing halos (bottom row), we compare the simulated central densities and core sizes reached at the time of minimum core density with the fluid-model prediction. Halos m10b,f are much farther from the model prediction, with $\rho_c/\rho_{s,\rm CDM} = 0.4$, $0.5$ and $r_{\rm core}/r_{\rm s,CDM} = 1.0$, $1.0$, respectively. Their central densities are therefore lower by a factor of $5.6$ and $4.5$, while their cores are larger by a factor of $2.2$ and $2.2$, relative to the fluid-model expectation. 
Interestingly, the m10b,f pair reach similar core density and radii despite having different merger histories---the same is true for the m10q,d pair. A larger sample size is needed to determine whether this similarity is a mere coincidence or instead reflects a new steady state reached by these halos. 
Comparatively, m10j reaches $\rho_{\rm c,0}/\rho_{\rm s, CDM} = 1.6$ and $r_{\rm core,0}/r_{\rm s, CDM} = 0.5$. Its central density is therefore lower by a factor of 1.2, while its core is larger by a factor of 1.5, relative to the fluid-model prediction.
Although this halo is not deep in the collapse regime, it goes beyond the point of minimum core density and begins to enter the collapse regime where the density increases and the radius decreases. 

The core densities reached by the collapsing and non-collapsing halos span three orders of magnitude from as high as $\sim 10^{10}~{\rm M}_{\rm \odot}/\rm kpc^3$ to as low as $\sim 10^{7}~{\rm M}_{\rm \odot}/\rm kpc^3$ (see Figure~\ref{fig:coreden}).\footnote{The exact core densities reached may be resolution dependent, however the mechanism for mergers to drive the decrease in core densities remains physical and resolution-independent, as discussed in Appendix~\ref{app:res}.} The low-density end is of particular interest because some of the halos reach densities a factor of $\sim5$ lower than predicted by the fluid model. Given the self-similar nature of the fluid model, these values are not possible to reach by varying the halos' initial conditions. Mergers in SIDM halos thus provide a novel pathway to achieve these low densities, one with no CDM analog, and may offer a new mechanism for creating gas-rich, DM-deficient dwarf galaxies in the field~\citep{Kong_2022}.

Figure~\ref{fig:analyticrhocrc} further provides the time it takes each halo to reach $\rho_{\rm c,0}$ in the fluid model, $t_{\rm core, mod}$, and in the simulations, $t_{\rm core, sim}$. For halos m10q,d,i, $t_{\rm core,sim}=147$, $264$, and $344\,t_{\rm c,0}$, compared with the $50\,t_{\rm c,0}$ predicted by the fluid model. Thus, the fluid model underestimates the time to maximal core by a factor of roughly $3$--$7$.\footnote{This discrepancy is still notable when accounting for uncertainties from $C$ and $t_{\rm core,mod}$. When these constants are tuned to decrease the difference between model and simulation (using $C=0.6$ and $t_{\rm core,mod} = 70 \, t_{\rm c,0}$), the fluid model still underestimates the time to maximal core by a factor of roughly $2$--$4$.}
Despite these differences, the best-fit logarithmic slopes of the cores of m10q,d,i in the core-expansion phase are comparable to those of the fluid model. The measured slopes,
% follow the fluid model with slope ratios, 
$\frac{{\rm d}\log \rho_{\rm c}/\rho_{\rm s, CDM}}{{\rm d}\log r_{\rm core}/r_{\rm s, CDM}}$, are $0.93$, $1.22$, and $1.28$ times the fluid-model value, respectively.

Next, we compare the simulations with the fluid model after core expansion during the gravothermal-collapse regime with the results summarized in Table~\ref{tab:tcoll}. To study the gravothermal-collapse regime independently of core expansion, the collapse time is measured as the interval between $t_{\rm core}$ and $t_{\rm coll}$, 
\begin{equation}
    \Delta t_{\rm coll} = t_{\rm coll} - t_{\rm core}\, .
    \label{eq:deltatcoll}
\end{equation}
The simulated collapse time interval, $\Delta t_{\rm coll, sim}$, is calculated using $t_{\rm core,sim}$ and $t_{\rm coll, sim}$, as defined in Section~\ref{sec:core_mergers}.
The predicted collapse time for the fluid model is $\Delta t_{\rm coll, mod} = t_{\rm coll,mod}-t_{\rm core,mod}=(400-50)\,t_{\rm c,0}=350\,t_{\rm c,0}$. 

The modeled collapse time is evaluated for two different methods.
The first is a CDM-based estimate, in which $t_{\rm c,0}$ is evaluated using the NFW parameters $\rho_{\rm s, CDM}$ and $V_{\rm max}$ of the corresponding CDM simulation at $z=0$. The predicted scaling relations introduced in Section~\ref{sec:fluidmodel} are then used to calculate $\rho_{\rm c,0}$ and $v_{\rm c,0}$.
Because Figure~\ref{fig:analyticrhocrc} shows that the simulated halos deviate from these predicted scaling relations, we also consider a second approach in which $t_{\rm c,0}$ is calculated from the core properties at the time of minimum density using the simulation-measured values of $\rho_{\rm c,0}$ and $v_{\rm c,0}$. This method provides a first attempt to model the gravothermal-collapse regime despite the discrepancies found above. 
% If a halo does not collapse in the simulation, the corresponding collapse time interval is omitted.

For the halos that collapse, $\Delta t_{\rm coll, mod}^{\rm CDM}$ systematically under-predicts $\Delta t_{\rm coll, sim}$, with fractional differences of $\sim 20\%$ for m10q,d and $\sim 40\%$ for m10i.\footnote{This discrepancy is still notable when accounting for uncertainties from $C$; the discrepancy remains at  $\sim5\%$ and $\sim25\%$ for m10q,d and m10i, respectively, when a value of $C=0.6$ is used.} 
These collapse times may be dependent on simulation mass resolution. Appendix~\ref{app:res} shows that lower-resolution runs of m10i, which do not meet the resolution criteria for idealized simulations~\citep{palubski2024numerical, Mace2024}, have not collapsed by the time the fiducial-resolution run does.

Despite the discrepancy between simulations and the fluid model in collapse time---and in addition to the discrepancies in the normalization of core density and radius discussed earlier---the post-expansion evolution of halos m10q,d,i closely follows the fluid model. The best-fit logarithmic slopes, $\frac{{\rm d}\log \rho_{\rm c}/\rho_{\rm s, CDM}}{{\rm d}\log r_{\rm core}/r_{\rm s, CDM}}$ of m10q,d,i are $0.99$, $1.01$, and $1.01$ times the fluid-model value, respectively. This indicates that while the fluid model does not accurately predict when collapse occurs, it robustly captures the scaling between core radius and core density. This result emphasizes that the density profile for halos which undergo core collapse can still be modeled using analytic profiles.

\begin{table}[t]
    % \centering
    \setlength{\tabcolsep}{0pt}
    \begin{tabular}{c c c c}
        \hline
        \textbf{Name} & \bm{$\Delta t_{\rm coll, sim}$} & \bm{$\Delta t_{\rm coll, mod}^{\rm CDM}$} & \bm{$\Delta t_{\rm coll, mod}^{\rm SIDM}$}\\
         & [Gyr] & [Gyr] & [Gyr] \\
        \hline
        m10q-s70 & 7.7 & 6.4 & 13.0 \\
        m10d-s70 & 5.9 & 4.7 & 8.8 \\
        m10i-s70 & 3.0 & 2.0 & 2.5 \\
        m10b-s70 & -& 8.9 & 47.7 \\
        m10f-s70 &-& 6.2 & 26.7 \\
        m10j-s70 &-& 2.0 & 3.0 \\
        \hline
    \end{tabular}
    \caption{Simulated and predicted gravothermal-collapse time intervals. The columns list: (1)~Halo name; (2)~Time interval between minimum core density and gravothermal collapse in simulations, $\Delta t_{\rm coll, sim}=t_{\rm coll, sim}-t_{\rm core, sim}$; (3)~Collapse time interval, $\Delta t_{\rm coll, mod}^{\rm CDM} = t_{\rm coll,mod}^{\rm CDM}-t_{\rm core,mod}^{\rm CDM}$, defined as $(400-50)\,t_{\rm c,0} = 350 \,t_{\rm c,0}$ using the corresponding CDM halo NFW properties; and (4)~Collapse time interval, $\Delta t_{\rm coll, mod}^{\rm SIDM} = t_{\rm coll,mod}^{\rm SIDM} - t_{\rm core,mod}^{\rm SIDM}$, defined using the core properties at the time of minimum density, $\rho_{\rm c,0}$ and $v_{\rm c,0}$.
    }
    \label{tab:tcoll}
\end{table}

Compared to $\Delta t_{\rm coll, mod}^{\rm CDM}$, $\Delta t_{\rm coll, mod}^{\rm SIDM}$ increases for all the halos because the core density reaches smaller values than those assumed in the scaling relation (Figure~\ref{fig:analyticrhocrc}), which directly increases the predicted collapse time. Table~\ref{tab:tcoll} shows that $\Delta t_{\rm coll, mod}^{\rm CDM}$ predicts $\Delta t_{\rm coll, sim}$ with smaller fractional differences than $\Delta t_{\rm coll, mod}^{\rm SIDM}$. We leave a detailed analysis of models that best predict cosmological simulations to future work. Similar limitations of the gravothermal fluid model in reproducing N-body results have been reported by~\citet{Jiang_2023}.

These results caution against using the fluid model to estimate the timescale for core collapse based on $\rho_{\rm s, CDM}$ and $r_{\rm s, CDM}$ (or equivalently the concentration and mass) of the corresponding CDM halo, except as a lower bound. Predictions of halos being in the core-expansion or -collapse phase for a given cross section are dependent on the merger history of the halo. The lack of knowledge of the merger history of a halo introduces a large uncertainty in this time scale, in the direction of making it longer than the analytic prediction. This directly translates into an uncertainty in the fraction of collapsed halos, although a larger suite of simulations is required to quantify this fraction. 
% At the same time, our results indicate that for s70, half of the simulated halos do enter the core-collapse regime,so the estimates using the fluid model are still useful. 
Nevertheless, once a halo gets into the core-collapse regime, the core-density--core-radius relation tracks the analytic expectation, suggesting that analytic halo density profiles would still be a reasonable approximation.

\section{Conclusions}
\label{sec:conclusion}

We study the impact of mergers on a halo's gravothermal evolution using a suite of cosmological DMO simulations with an SIDM cross section of $70$~\cmg{} that are predicted to collapse within the age of the Universe. We find that mergers can inject orbital kinetic energy into the halo, altering the heat transport in some cases such that core collapse is delayed and central densities are driven below the expectation from isolated evolution. 
The underlying reason for this is that, at these large cross sections, the probability for interaction between subhalo and halo particles is order unity if the subhalo gets close to the core of the main halo.  

Our main findings are as follows:

\begin{itemize}
    \item Although all halos in this work are expected to undergo gravothermal collapse within a Hubble time based on the fluid model~\citep{Outmezguine_2023}, only half of them (m10q,d,i) reach deep into the gravothermal-collapse regime (i.e., a final core density of at least $10$ times the minimum core density), while the other half (m10b,f,j) do not. This introduces an uncertainty in predictions for the gravothermal collapse time scale of field halos in SIDM models with a high cross section, with the analytic prediction providing a lower bound. Additionally, it increases the diversity of the SIDM predictions beyond that predicted from the concentration-mass relation coupled with the gravothermal evolution~\citep{Roberts:2024uyw}. 

    \item To diagnose the physics behind the impact of mergers on gravothermal evolution, we study the change in heat flow. The average temperature within $1$~kpc, $\widetilde{T}_{\rm inner}$, is compared with that measured over the radial range $1$--$5$~kpc, $\widetilde{T}_{\rm outer}$. The halos that do not collapse experience episodes of inward heat flow, each correlated with subhalo infall. 

    \item The primary causes that impact halo evolution are the merger mass ratio, radiality, and timing. %(later mergers beyond the core-expansion phase but before the halo is deep in the core-collapse regime are more likely to stall gravothermal collapse). 
    We propose that the maximum quiescence time interval between mergers, $\Delta t_{\rm max}/t_{\rm relax}$, is another useful diagnostic for determining whether a halo will collapse. For example, m10b,f have  $\Delta t_{\rm max}/t_{\rm relax} < 100$, while m10q,d,i,j have $\Delta t_{\rm max}/t_{\rm relax}>250$.  Halos with mergers that are too frequent---specifically, less than the timescale needed for an isolated halo to reach maximal core---may not collapse. 
    
    \item The halos that undergo gravothermal collapse do not consistently follow the fluid-model predictions during the core-expansion phase. All halos reach maximum cores with larger radii and lower densities than predicted. Furthermore, the fluid model underestimates the time to core expansion by a factor of roughly $2$--$6$. 

    \item 
    Estimating the time to gravothermal collapse based on the concentration and mass of the corresponding CDM halo is not reliable, as it can be significantly longer. The estimates work best for halos that are far along the core-collapse regime by the end of the simulations, with the time to gravothermal collapse being about $20$--$40\%$ longer than predicted from the corresponding CDM halo. 
    
    \item 
    The central densities in two halos that do not core collapse (m10b,f) reach values much lower than what can be achieved in the gravothermal model ($\sim2.4\,\rho_{\rm s, CDM}$). The interplay of mergers and heat transport lead to both the lowering of central densities and the delay in core collapse. Our results suggest that the central halo densities in SIDM models with large cross sections have a wider range than predicted previously, and they motivate higher resolution N-body as well as hydrodynamic simulations to explore whether this could be a new pathway for creating gas-rich dwarf galaxies in the field that are deficient in DM~\citep{Kong_2022}. 
  
\end{itemize}

These results demonstrate the importance of a halo's merger history on its predicted gravothermal evolution. 
They should be confirmed with future studies run on a larger sample of initial conditions, with even higher resolution, and including baryonic physics. The combination of adiabatic contraction due to baryons~\citep[e.g.,][]{despali_aida-tng_2025}, feedback-induced cores~\citep[e.g.,][]{mashchenko_stellar_2008}, and SIDM gravothermal evolution is non-trivial and requires full hydrodynamic simulations to determine how these processes jointly shape the diversity of dwarf galaxy density and circular velocity profiles.

Building a model that describes the direct influence between mergers and gravothermal evolution will also require supplementing with idealized simulations that systematically vary merger properties---such as orbital parameters, mass ratios, and infall times---in a controlled manner. We note that halos farther along in the core-collapse regime do track the core-density--radius relation of the fluid model reasonably well. This indicates that we can still use analytic halo profiles deep in the core-collapse regime. However, predictions of a halo's stage in the gravothermal evolution from its core density and radius must additionally take into account decreased core densities and increased core radii due to mergers.

The endeavor to build a model that accounts for mergers is motivated by our key result that the interplay of mergers and heat transport lead to significant delays for some halos in getting to the core-collapse regime with an associated decrease in the core densities. 
In the absence of a predictive framework to identify which halos can get to the core-collapse regime, our work highlights the importance of marginalizing over the halo environment, in particular merger history, when predicting the gravothermal evolution of field halos.

\begin{contribution}
MS generated initial conditions, executed and coordinated simulation runs, developed analysis code, performed the analysis, and drafted the manuscript.
AH implemented major revisions to the simulation code by updating the time-stepping procedure, generated initial conditions, executed simulation runs, developed analysis metrics, and contributed to writing the manuscript.
AA executed simulation runs and interpreted the results.
ML acquired funding and computation resources and oversaw the project.
LN acquired funding and computation resources, oversaw the project, and provided interpretation in the context of the FIRE Collaboration.
MK oversaw the project.
AT executed simulation runs.
SO interpreted the results.
RS provided interpretation in the context of the FIRE Collaboration.
XS provided interpretation in the context of the FIRE Collaboration.
JM provided interpretation in the context of the FIRE Collaboration and puzzles in the low-mass regime.
All authors reviewed and edited the manuscript.
\end{contribution}

\begin{acknowledgments}

We thank Akaxia Cruz, Sophia Gad-Nasr, Ethan Nadler, Andrew Robertson, Oren Slone, Frank van den Bosch, and Hai-bo Yu for insightful discussions and comments. We are also grateful to the Scientific Computing Team at the Flatiron Institute for computing resources and support that shaped this work. MS was supported by a research grant (VIL53081) from VILLUM FONDEN. 
AH is supported by NSF award 2307788. 
AA is supported by the Gordon and Betty Moore Foundation. 
ML is supported by the National Science Foundation~(NSF), under Award Number AST~2307789, as well as the Simons Investigator in Physics Award.   
LN is supported by NSF awards 2307788, CAREER award AST-2337864 and the Sloan fellowship. JM is funded by a Pomona College Large Research Grant.
This work was performed in part at Aspen Center for Physics, which is supported by National Science Foundation grant PHY-2210452. Simulations were run on the Texas Advanced Computing Center (TACC; \url{https://tacc.utexas.edu/}) and stored on the Princeton Research Computing TigerData system (\url{https://tigerdata.princeton.edu/}).

AH acknowledges that this manuscript has been authored by Fermi Forward Discovery Group, LLC under Contract No. 89243024CSC000002 with the U.S. Department of Energy, Office of Science, Office of High Energy Physics. AH acknowledges that this material is based upon work supported by the U.S. Department of Energy, Office of Science, Office of Workforce Development for Teachers and Scientists, Office of Science Graduate Student Research (SCGSR) program. The SCGSR program is administered by the Oak Ridge Institute for Science and Education (ORISE) for the DOE. ORISE is managed by ORAU under contract number DESC0014664. All opinions expressed in this paper are the author’s and do not necessarily reflect the policies and views of DOE, ORAU, or ORISE. AH would like to thank the Kavli Institute for Cosmological Physics~(KICP) at the University of Chicago for their hospitality, where part of this work was conducted. 

MS used ChatGPT (OpenAI) in the development of this work. ChatGPT was used to develop the python code to create figures and debug error messages. It was occasionally used to offer ideas for rephrasing sentences that explain complex concepts to improve the clarity of this paper. In the context of code development, ChatGPT was asked to edit visual aspects of figures (e.g., color schemes, notation placement, line style specification, etc.) and write comments for individual functions. In the context of debugging, ChatGPT was asked to summarize error messages, explain possible sources for the error, and offer suggestions for solving the bug. In the context of editing, ChatGPT was asked to offer options for rephrasing a sentence in a clear and academic tone. MS carefully reviewed all of ChatGPT's outputs and takes full responsibility for all code used in this work. All scientific conclusions are the authors’ own.

\end{acknowledgments}

\software{This work made use of the following software packages:
        \texttt{Gizmo} \citep{hopkins15}
        \texttt{python} \citep{python},
        \texttt{halo analysis}~\citep{Wetzel:2016apj},
        \texttt{gizmo analysis}~\citep{Wetzel:2016apj},
        \texttt{pandas}~\citep{mckinney2011pandas},
        \texttt{numpy}~\citep{harris2020array},
        \texttt{matplotlib}~\citep{Hunter:2007},
        \texttt{scipy}~\citep{2020SciPy-NMeth},
        \texttt{astropy}~\citep{astropy:2013},
        \texttt{colossus}~\citep{Diemer_2018}.
}

\counterwithin{figure}{section}
\counterwithin{table}{section}

\appendix
\section{Adaptive Time-Stepping Parameters}
\label{app:timestep}

As discussed in Section~\ref{sec:sidm_imp}, simulation time-stepping can affect halo evolution, particularly when including non-gravitational self interactions~\citep{van_den_Bosch_2018, Hopkins_2018, palubski2024numerical, Mace2024}. An adaptive mechanism allows the timestep for each particle to be calculated based on its local density and gravitational acceleration, such that it is small enough to resolve physical interactions while  not wasting resources on particles with infrequent interactions. In this Appendix, we vary two numerical parameters, $\kappa$ and $\eta$, which set the adaptive timesteps, to determine values that lead to a converged gravothermal evolution. 

As defined in Equation~\ref{eq:probij}, $\kappa$ determines the upper limit for the probability that a particle scatters with any of its neighbors within a single timestep. This parameter is designed to restrict the number of multiple scatters in a single timestep, which may lead to energy and momentum non-conservation~\citep{palubski2024numerical}.  
The size of the timestep is set by the gravitational acceleration that a particle experiences within $\delta t_i$. The tolerance parameter, $\eta$, defined in Equation~\ref{eq:eta}, determines how strongly the timestep scales inversely with gravitational acceleration.
$\eta$ becomes important in high-density regions where particles experience large accelerations. 
The default values for $\kappa$ and $\eta$ in \texttt{GIZMO} are $0.2$ and $0.02$, respectively. However, studies of gravothermal evolution in isolated non-cosmological halos find that more constrained values can reduce artificial heating during the collapse process~\citep{palubski2024numerical, Mace2024}. 

To identify the optimal values for $\kappa$ and $\eta$ for our cosmological N-body SIDM simulations, we systematically scan over different combinations, probing smaller values than explored in previous works. Specifically, we scan $\kappa$ and $\eta$ values of $0.01$, $0.002$, and $0.0005$ using m10q because of its relatively quiet merger history and clear gravothermal evolution.

\begin{figure}[ht]
    \centering
    \includegraphics[width=0.5\textwidth]{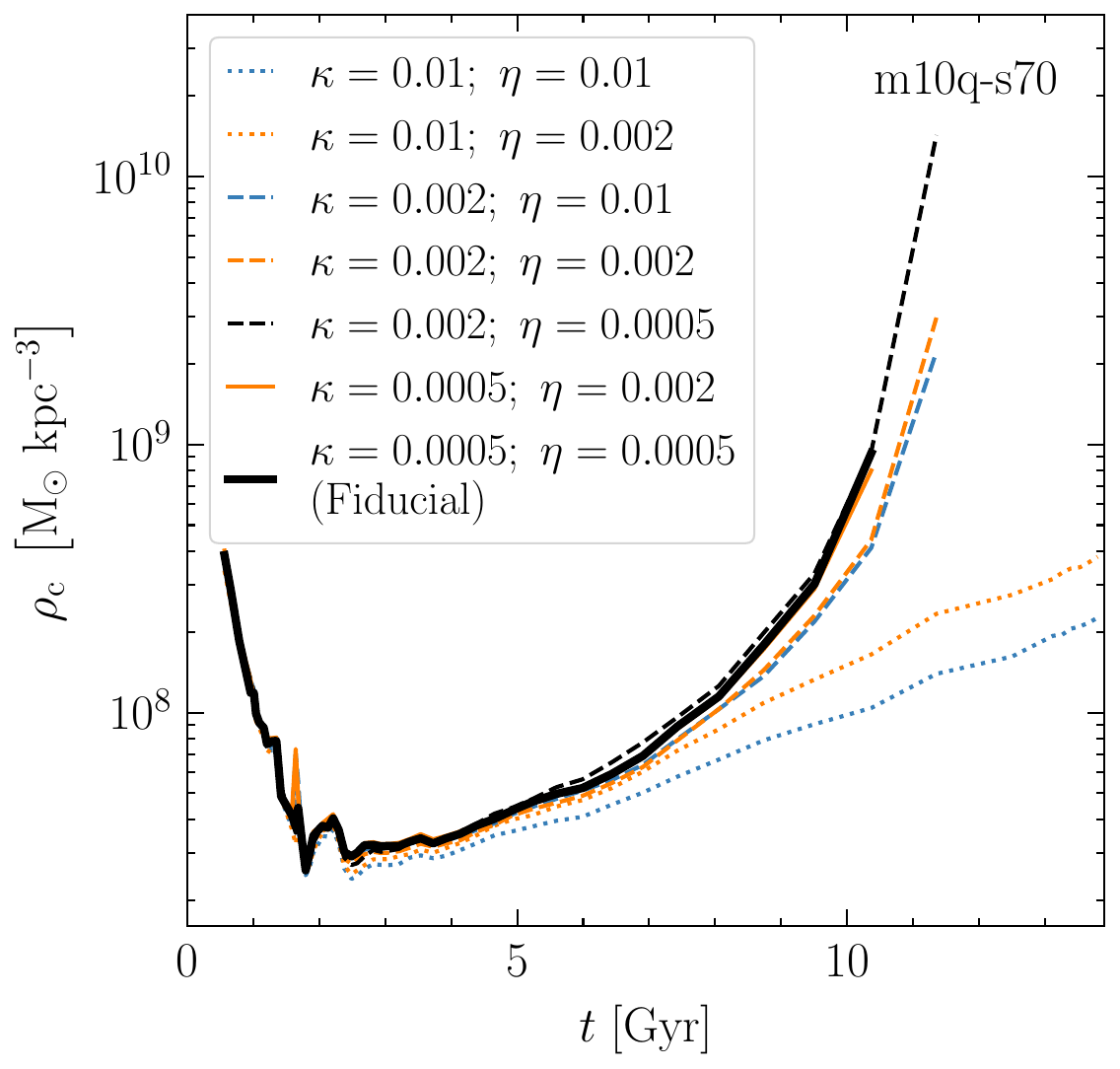}
    \qquad
    \caption{Evolution of m10q-s70 core densities with time-stepping parameters, $\kappa$ and $\eta$, varied. $\eta$ values are $0.01$ (blue), $0.002$ (orange), and $0.0005$ (fiducial; black). $\kappa$ values are $0.01$ (dotted), $0.002$ (dashed), and $0.0005$ (fiducial; solid). To highlight the fiducial values used throughout this paper, $\kappa = 0.0005$ and $\eta = 0.0005$ is plotted with a thicker line. The lowest values of $\kappa$ and $\eta$ lead to abnormally low core densities and long collapse times. By $\kappa = 0.0005$ and $\eta=0.0005$, the evolution becomes insensitive to the choice of these parameters, indicating convergence.
    } \label{fig:kescan}
\end{figure}

Figure~\ref{fig:kescan} shows how the core density of m10q evolves with time, for a cross section of 70~\cmg{} and different combinations of $\kappa$ and $\eta$. The core density is defined in Section~\ref{sec:properties}. 
At lower $\kappa$ or $\eta$ values, the core density reaches lower minimums and later collapse times. In the most extreme case, the $\kappa=0.01$ and $\eta=0.01$ run reaches a minimum core density $0.83$ times the minimum core density in the $\kappa=0.0005$ and $\eta=0.0005$ run. 

The runs with $\kappa = 0.002$ and $\eta = 0.0005$, $\kappa = 0.0005$ and $\eta = 0.002$, and $\kappa = 0.0005$ and $\eta = 0.0005$ are converged. Since these three runs use similar computational resources, we chose to use the most conservative values; 
% Convergence occurs at $\kappa = 0.0005$ and $\eta = 0.0005$. 
we therefore adopt $\kappa = 0.0005$ and $\eta = 0.0005$ as the fiducial values for all the simulations discussed in this work. While these values are converged for halo m10q-s70, they are not guaranteed to be optimal for all ICs. Due to computational expense, we postpone such an exploration for future work. 

As evident from Figure~\ref{fig:kescan}, not all simulations are run to completion. This is because particles with extremely small timesteps increase the simulation cost.
We terminate a run before $z=0$ if it does not reach the next snapshot within $1.1 \times 10^4$~CPU hours; the snapshot intervals range from $300$--$770$~Myr, depending on the termination time.

\section{Resolution Tests}
\label{app:res}

\begin{figure}
    \centering
    \includegraphics[width=0.6\linewidth]{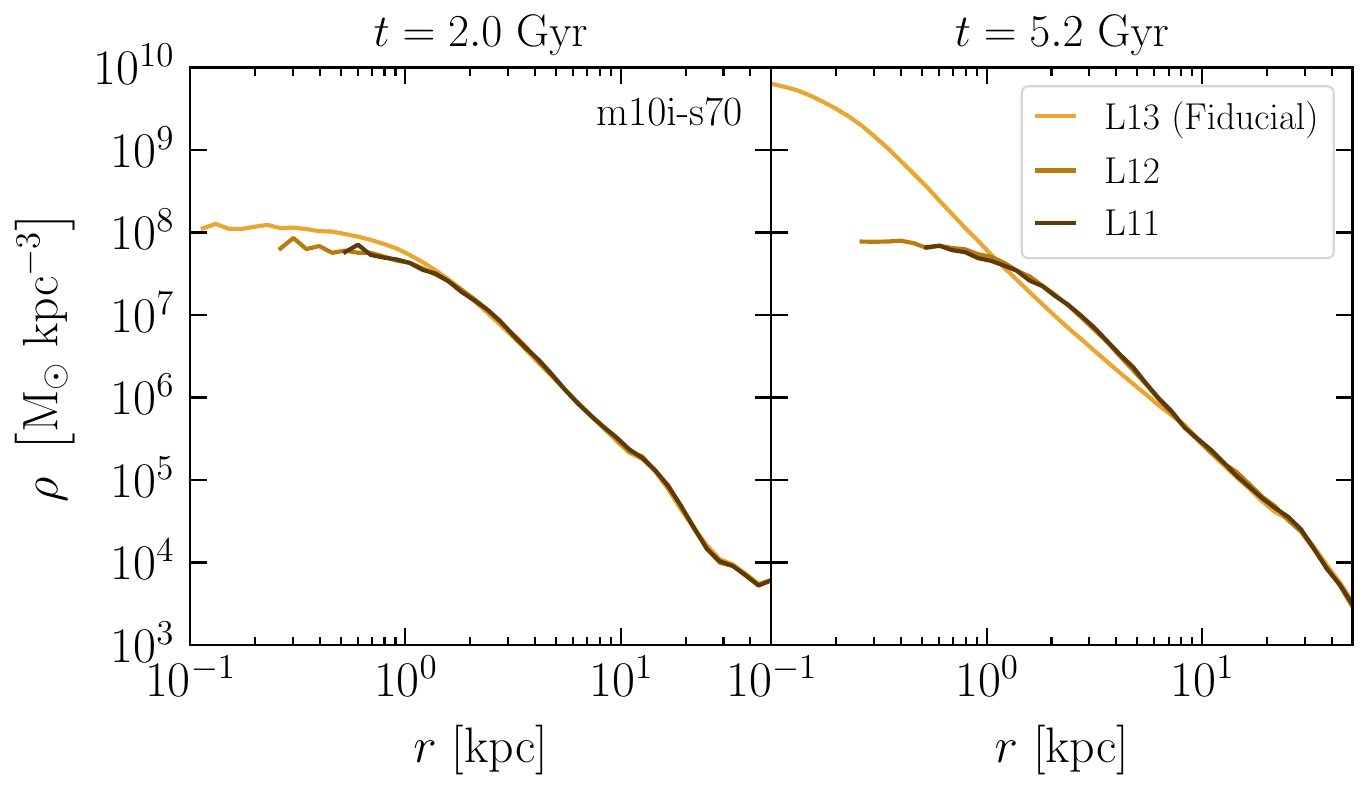}
    \includegraphics[width=0.6\linewidth]{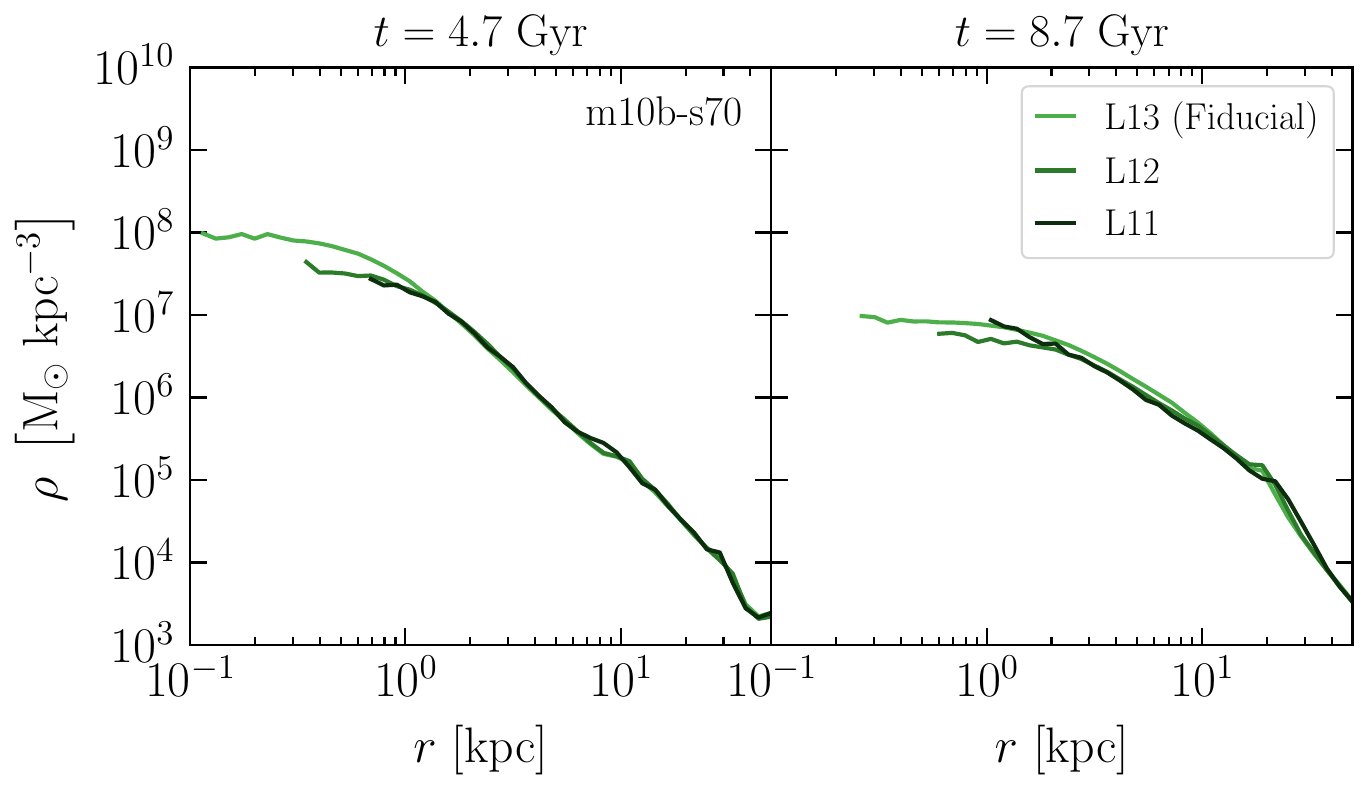}
    \caption{Density profiles of halos m10i-s70 (top) and m10b-s70 (bottom) at the fiducial resolution (L13), and at two lower resolutions with $8$ (L12) and $64$ (L11) times more massive DM particles. Density profiles are only plotted at radii that enclose at least $200$ particles. We show each halo at the latest time that all three resolutions are run to ($5.2$~Gyr for m10i and $8.7$~Gyr for m10b) and at a time that shows the early halo evolution, which for m10i-s70 corresponds to the time of minimum density~($2.0$~Gyr) and for m10b-s70 corresponds to a time before the major merger ($4.7$~Gyr). 
    When in the gravothermal-collapse regime, higher-resolution simulations collapse faster.
    }
    \label{fig:resolution_testing}
\end{figure}

This appendix explores how numerical resolution may affect the simulations presented here. Previous works have developed resolution requirements to accurately simulate cosmological CDM~\citep{Power03, van_den_Bosch_2018, Hopkins_2018, Wheeler_2019} and idealized SIDM simulations~\citep{palubski2024numerical, Mace2024}.
For CDM DMO simulations, numerical convergence is expected beyond the Power radius~\citep{Power03}. When applied to CDM hydrodynamic simulations of isolated $M_{\rm host} = 10^{10}$~\Msolar{} halos,~\citet{Fitts2019} find that the Power radius, corresponding to the radius that encloses $\sim2500$ particles, conservatively indicates the point of numerical convergence. 
Furthermore,~\citet{Wheeler_2019} find that CDM hydrodynamic simulations of isolated dwarf galaxy halos at the standard FIRE resolution ($m_{\rm \scriptscriptstyle DM} = 1.5\times 10^3$~\Msolar), which is about the fiducial resolution used in this work, are consistent with simulations with extremely high resolution ($m_{\rm \scriptscriptstyle DM} = 150$~\Msolar) beyond the Power radius.

Because SIDM causes some halos to evolve into the gravothermal-collapse regime, where the core reaches higher densities than in CDM, these convergence requirements may not be sufficient for SIDM. In idealized simulations of SIDM dwarf galaxy halos, $10^6$ DM particles within the halo are required for halos to consistently undergo gravothermal collapse with $3\%$ scatter in collapse times across different realizations of the same halo~\citep{Mace2024} and $10\%$ scatter in minimum core densities~\citep{palubski2024numerical}. Halos with fewer particles have a greater scatter in both collapse times and minimum core densities.

This appendix considers simulations run at one and two steps lower mass resolution than the fiducial ones studied in this work. 
Figure~\ref{fig:resolution_testing} shows the density profiles for three different cases: (1)~L13, or fiducial, with $m_{\rm \scriptscriptstyle DM} = 3\times 10^3$~\Msolar{} and $\epsilon_{\rm \scriptscriptstyle DM} = 28$~pc, (2)~L12 with $m_{\rm \scriptscriptstyle DM} = 2.4\times 10^4$~\Msolar{} and $\epsilon_{\rm \scriptscriptstyle DM} = 56$~pc, and (3)~L11 with $m_{\rm \scriptscriptstyle DM} = 1.9\times 10^5$~\Msolar{} and $\epsilon_{\rm \scriptscriptstyle DM} = 110$~pc. All simulations presented throughout this work enclose at least $3\times 10^6$ particles within their virial radii, while the L12 and L11 simulations presented in this appendix enclose $\sim 4 \times 10^5$ and $5\times 10^4$ particles, respectively.

To bracket the effects, we explore the impact of mass resolution on m10i-70, which reaches the highest core density due to gravothermal collapse, and m10b-s70, which reaches the lowest core density due to a major merger. Figure~\ref{fig:resolution_testing} shows the density profiles for halo m10i-s70 at the time of minimum density~($2.0$~Gyr) and at the last available snapshot~($5.2$~Gyr). For halo m10b-s70, the density profiles are shown for a time before the major merger~($4.7$~Gyr) and at the last available snapshot~($8.7$~Gyr). 

For halo m10i-s70, the L11 and L12 runs agree at both the early and late time.
However, the L13 run has a larger core density within $\sim1$~kpc than the L11 and L12 runs. At $2.0$~Gyr,
this is due to numerical heating driving the low-resolution simulations to lower minimum densities~\citep{palubski2024numerical, Mace2024}.
At $5.2$~Gyr, the L13 run has a core density almost 100 times that of the L12 and L11 runs. This is because the L13 run is already deep in the gravothermal-collapse regime (shown in Figure~\ref{fig:coreden}) while the L12 and L11 runs experience a delayed collapse time. As discussed in ~\citet{palubski2024numerical, Mace2024} an increase in numerical resolution can decrease the collapse time, until the point that convergence is reached.  

The results for halo m10b-s70 are similar, although less dramatic, since it never reaches deep within the collapse regime. In particular, the L13 run has a larger core density than the L12 and L11 runs at $4.7$~Gyr. This is due to the L13 run undergoing early gravothermal collapse before the major merger (shown in Figure~\ref{fig:coreden}) while the L12 and L11 runs do not experience this early collapse due to numerical heating. At $8.7$~Gyr (after the major merger), all three resolutions give roughly the same results.

%This resolution test shows that the fiducial runs (L13) are consistent with the lower-resolution runs only when the halo is not undergoing gravothermal collapse. The L13 runs undergo gravothermal collapse faster than their low-resolution counterparts, resulting in larger core densities. This speed-up in collapse time may be due to increased scatter in the generation of the ICs, as hypothesized in~\citet{palubski2024numerical}. 

While the halos run at the fiducial L13 resolution satisfy the convergence criteria derived from isolated SIDM simulations~\citep{palubski2024numerical, Mace2024}, robustly demonstrating their convergence in a cosmological setting would require rerunning the simulations at one-level higher in mass resolution.  Unfortunately, the computational cost of this is prohibitive. A CDM simulation at the next higher resolution would be expected to require 16 times the computational resources of the fiducial resolution run, while an SIDM run would require 1.3 times the computational resources of a CDM run.
Taken together, this implies a minimum computational cost of about 21 times the fiducial cost for a higher-resolution SIDM run. We note throughout the main body where our results may be affected by resolution.

\section{Core Radius}
\label{app:corerad}

\begin{figure}
    \centering
    \includegraphics[width=0.39\linewidth]{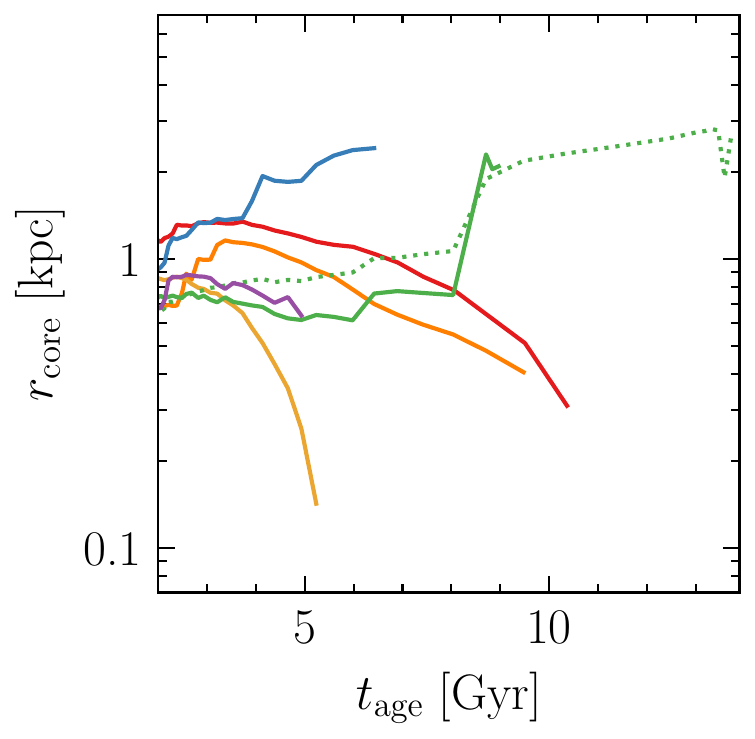}
    \qquad
    \includegraphics[width=0.41\linewidth]{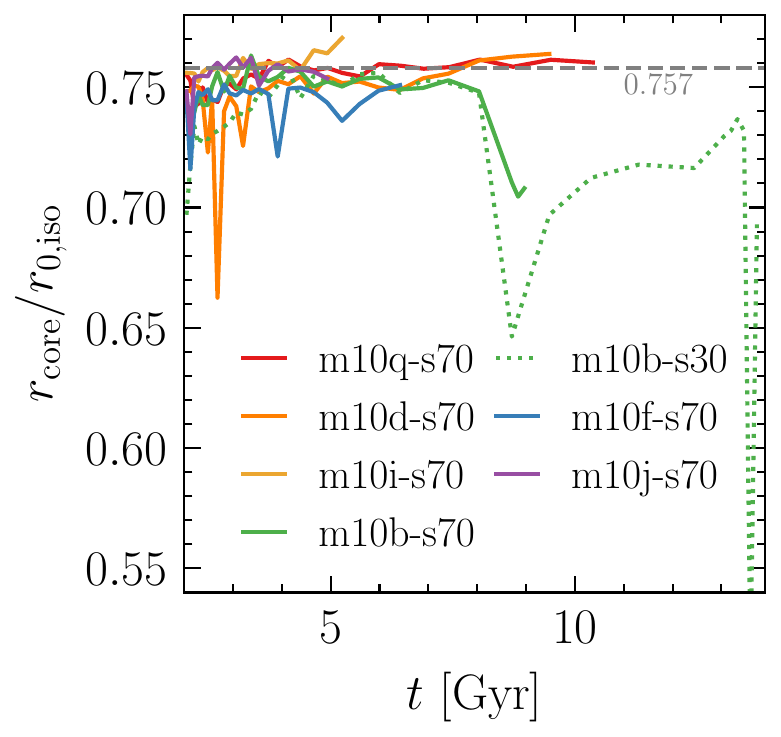}
    \caption{\textit{(Left)}~Core radius of each host halo (m10q,d,i,b,f,j) with $\sigma/m=70$~\cmg{}, as well as m10b-s30, across time. Core radius is defined as the radius at which the fit analytic density falls to $1/2$ the central density. Halos that undergo gravothermal collapse (m10q,d,i) have rapidly decreasing $r_{\rm core}$ at late times, while those that don't (m10b,f) have an increasing $r_{\rm core}$. m10j is in the early stages of collapse and has a slightly decreasing $r_{\rm core}$.
    \textit{(Right)}~Core radius normalized by $r_{\rm 0, iso}$ of each host halo with $\sigma/m=70$~\cmg{}, as well as m10b-s30, across time. Halos with isothermal cores follow the predicted relationship, $r_{\rm core}/r_{\rm 0, iso} = 0.757$ (dashed gray), while halos with major or frequent mergers have smaller normalized core radii. Radii are only shown after $2$~Gyr since early halos are not expected to form isothermal cores.
    }
    \label{fig:corerad}
\end{figure}

This appendix discusses how the core radius, $r_{\rm core}$, of each host halo in the suite evolves with time. The left panel of Figure~\ref{fig:corerad} plots the time dependence of $r_{\rm core}$ for the six s70 halos, as well as m10b-s30. $r_{\rm core}$ is defined as the radius where the density falls to $1/2$ its central value, obtained by fitting the analytic profile in Equation~\ref{eq:coreden} to each host. Consistent with Figure~\ref{fig:coreden}, halos m10q,d,i undergo gravothermal collapse, and m10j is entering the gravothermal-collapse regime: in each case, $r_{\rm core}$ initially increases, but then turns around and decreases. In contrast, m10b,f do not undergo gravothermal collapse and have $r_{\rm core}$ that increases most of the time. Halo m10b-s30 also does not collapse and thus has an increasing $r_{\rm core}$.

Section~\ref{sec:mass_energy} showed that mergers impact the difference in the average temperature in the inner and outer halo. 
To define this difference, we adopted fixed radial ranges for the ``inner'' and ``outer'' regions. Specifically, we calculated the average temperature within $1$~kpc and compared it to that between $1$--$5$~kpc. As seen in the left panel of Figure~\ref{fig:corerad}, the inner radius encloses the core for most of the halo's evolution, except for halo m10f, which reaches a maximum core radius of $2.4$~kpc, the largest among the sample. The outer radius extends beyond twice this value, ensuring the outer region always includes particles outside the core.

The right panel of Figure~\ref{fig:corerad} shows the core radius normalized by the scale radius, which is useful for comparing to the isothermal profile,
\begin{equation}
    r_{\rm 0, iso} = \frac{3 v_{\rm c}}{\sqrt{4 \pi G \rho_{\rm c}}},
    \label{eq:r0}
\end{equation}
where $v_{\rm c}$ is the one-dimensional velocity dispersion within the core radius and $\rho_{\rm c}$ is the core density.
The isothermal model predicts that $r_{\rm core}/r_{\rm 0, iso} = 0.757$.
By comparing simulations to the isothermal prediction, we can test whether they have isothermal profiles in the core.  As evident from Figure~\ref{fig:corerad}, most of the s70 halos have core radii consistent with isothermal.  Notably, however, m10b,f fall below the predicted ratio.  These two halos experience major and frequent mergers that inject energy into the core, as shown in Figure~\ref{fig:coreenergy}. In the case of m10b-s30, the normalized $r_{\rm core}$ remains below the isothermal value until $z=0$.

\section{Testing Different SIDM Models}\label{app:diffSSIDM}

Given the dramatic density change it undergoes during its last major merger, m10b provides a useful case study to investigate how different cross sections impact heat flow and core evolution. Different cross sections can test the  hypothesis that mergers inject energy into the halo via self interactions. The extent to which a merger event impacts the core density evolution of an SIDM halo depends on the efficiency of heat flow between the merging subhalo's pericenter and the host core. More efficient heat flow allows for energy to be transferred from the merging subhalo to the host core, impacting the velocity dispersion and the density of the core. 

\begin{figure}[t]
    \centering
    \includegraphics[width=0.8\linewidth]{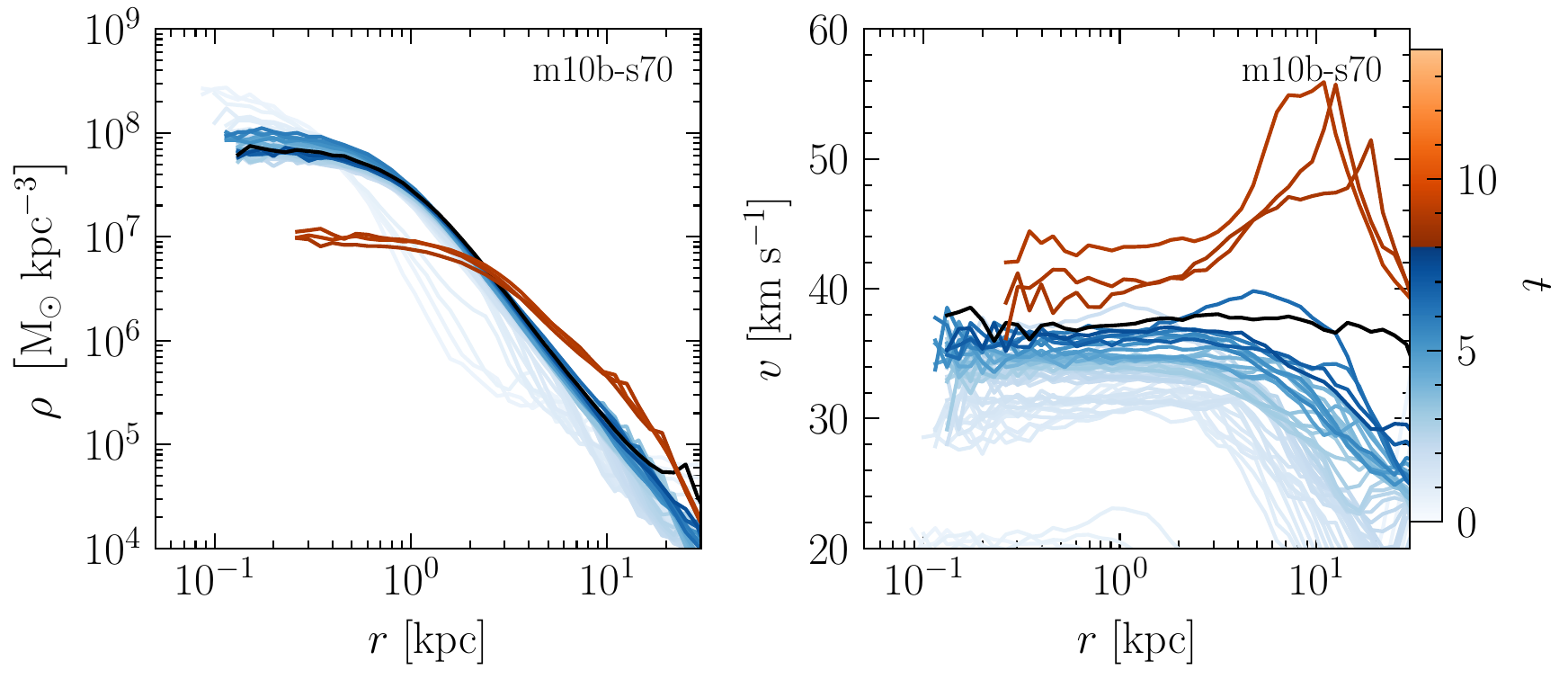}\\
    \includegraphics[width=0.8\linewidth]{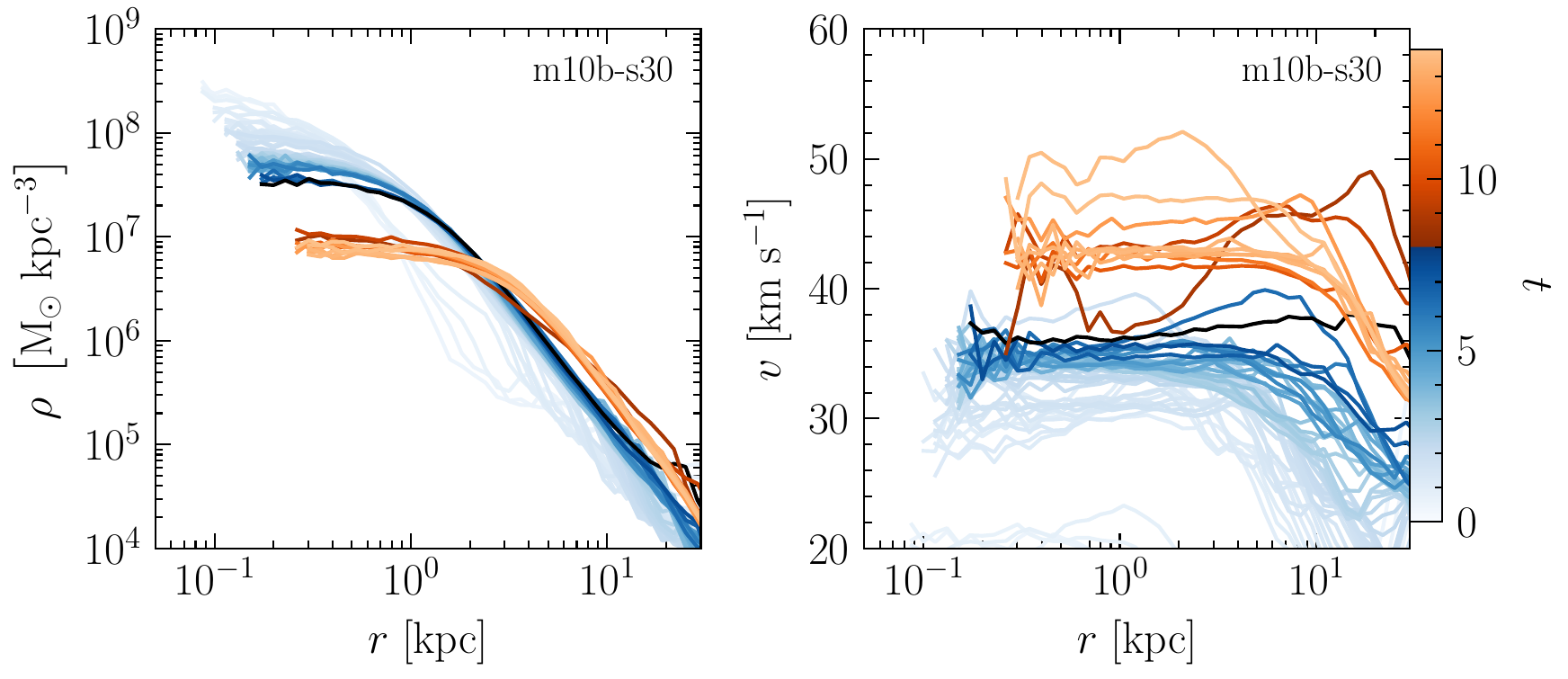}\\
    \caption{Density (left) and one-dimensional velocity dispersion (right) profiles for m10b with two SIDM models: $\sigma/m =70$~\cmg (top) and $\sigma/m =30$~\cmg (bottom). Halos in both models experience a major merger at $7.7$~Gyr. 
    The profiles are colored by their timing relative to the merger: blue indicates pre-merger (with lighter shades at earlier times), black marks the time of the merger infall, and orange indicates post-merger (with lighter shades at later times).
    }
    \label{fig:veldispm10b}
\end{figure}

In this appendix, we vary the self-interaction cross section, which varies the efficiency of heat flow, to study the impact of the radial major merger in m10b. Figure~\ref{fig:veldispm10b} shows the density and one-dimensional velocity dispersion profiles across time for halo m10b run with two SIDM models: $\sigma/m =70$~\cmg (top) and $\sigma/m =30$~\cmg (bottom). The profiles are colored by their timing relative to the merger: blue indicates pre-merger (with lighter shades at earlier times), black marks the time of the merger infall, and orange indicates post-merger (with lighter shades at later times). 

The s30 and s70 runs experience a $1:3$ merger at 7.7~Gyr, corresponding to a slightly positive velocity-dispersion slope from the center out to $\sim10$~kpc (black line). 
After the merger, both simulations develop a peak in velocity dispersion at $\sim10$~kpc.
In the s70 run, the peak remains at $\sim10$--$20$~kpc until the end of the simulation, while the core (within $2$~kpc; Figure~\ref{fig:corerad}) remains thermalized, albeit with a larger central velocity dispersion than before the merger.\footnote{Note that the final two snapshots of the m10b-s70 run are separated by shorter time intervals than those of m10b-s30 to follow the post-merger evolution as closely as possible.}
In contrast, m10b-s30 develops an isothermal core within $1.44$~Gyr of the merger infall, likewise with larger central velocity dispersion than before the merger.
By the end of the simulations, the velocity dispersions within $1$~kpc increase from $37$ to $43$~\kms and from $36$ to $47$~\kms for the s70 and s30 runs, respectively. This leads to a substantial decrease in core density following the merger. 

\clearpage
\pagebreak

\setlength{\bibsep}{0pt} % Reduces space between entries
\renewcommand{\baselinestretch}{1.0} % Remove paragraph spacing
\bibliography{biblio}{}
\bibliographystyle{aasjournalv7}

\end{document}